\documentclass[fleqn,usenatbib]{mnras}

\usepackage[T1]{fontenc}

\DeclareRobustCommand{\VAN}[3]{#2}
\let\VANthebibliography\thebibliography
\def\thebibliography{\DeclareRobustCommand{\VAN}[3]{##3}\VANthebibliography}

\usepackage{graphicx}	
\usepackage{amsmath}	
\usepackage{amssymb}	
\usepackage{bm}
\usepackage{mathtools}

\DeclarePairedDelimiter{\norm}{\lVert}{\rVert}
\newcommand\bigzero{\makebox(0,0){\text{\huge0}}}

\newcommand\rmi{\mathrm{i}}

\newcommand{\rhog}{{\rho_{\rm g}}}
\newcommand{\hatrhog}{{\hat{\rho}_{\rm g}}}
\newcommand{\rhogn}{{\rho_{\rm g}^0}}
\newcommand{\rhogo}{{\rho_{\rm g}^1}}
\newcommand{\vg}{{\bf{v}_{\rm g}}}
\newcommand{\vgo}{{\bf v}_{\rm g}^1}
\newcommand{\vge}{{\bf v}_{\rm g}^0}
\newcommand{\hatvg}{{\bf{\hat{v}}_{\rm g}}}

\newcommand{\vgx}{{v_{{\rm g}x}}}
\newcommand{\hatvgx}{{\hat{v}_{{\rm g}x}}}
\newcommand{\vgxn}{{v^0_{{\rm g}x}}}
\newcommand{\vgxo}{{v^1_{{\rm g}x}}}

\newcommand{\hatsigma}{{\hat{\sigma}}}

\newcommand{\rhod}{{\rho_{\rm d}}}
\newcommand{\rhodn}{{\rho_{\rm d}^0}}
\newcommand{\ud}{{\bf{u}}}

\newcommand{\uxn}{{u^0_{x}}}

\newcommand{\vd}{{\bf v}_{\rm d}}

\newcommand{\rmd }{{\rm d}}  

\newcommand{\taus}{{\tau_{\rm s}}}
\newcommand{\tausmin}{{\tau_{\rm s,min}}}
\newcommand{\tausmax}{{\tau_{\rm s,max}}}

\defcitealias{paper1}{Paper~I}
\defcitealias{paper2}{Paper~II}
\defcitealias{2005ApJ...620..459Y}{YG05}
\defcitealias{2019ApJ...878L..30K}{K+19}

\newcommand{\quadpack}{{\sc quadpack}}

\newcounter{foo}

\title[Polydisperse Streaming Instability II]{Polydisperse Streaming Instability II. Methods for solving the linear stability problem}

\author[S.-J. Paardekooper et al.]{
Sijme-Jan Paardekooper,$^{1,2}$
\thanks{E-mail: s.j.paardekooper@qmul.ac.uk,
 colin@colinmcnally.ca, f.lovascio@qmul.ac.uk}
\setcounter{foo}{\value{footnote}}
Colin P.~McNally,$^{1}$\footnotemark[\value{foo}]
\thanks{Present address: 81 Concession 8~E, Freelton, ON, L8B~1N9, Canada}
Francesco Lovascio$^{1}$\footnotemark[\value{foo}]
\\
$^{1}$Astronomy Unit, School of Physics and Astronomy, Queen Mary University of London, London E1 4NS, UK\\
$^{2}$DAMTP, University of Cambridge, Wilberforce Road, Cambridge CB3 0WA, UK
}

\date{Accepted XXX. Received YYY; in original form ZZZ}

\pubyear{2020}

\begin{document}
\label{firstpage}
\pagerange{\pageref{firstpage}--\pageref{lastpage}}
\maketitle

\begin{abstract}
Occurring in protoplanetary discs composed of dust and gas, streaming instabilities are a favoured mechanism to drive the formation of planetesimals.
The Polydispserse Streaming Instability is a generalisation of the Streaming Instability to a continuum of dust sizes.
This second paper in the series provides a more in-depth derivation of the governing equations and presents novel numerical methods for solving the associated linear stability problem.
In addition to the direct discretisation of the eigenproblem at second order introduced in the previous paper, a new technique based on numerically reducing the system of integral equations to a complex polynomial combined with root finding is found to yield accurate results at much lower computational cost.
A related method for counting roots of the dispersion relation inside a contour without locating those roots is also demonstrated.
Applications of these methods show they can reproduce and exceed the accuracy of previous results in the literature, and new benchmark results are provided.
Implementations of the methods described are made available in an accompanying Python package {\tt psitools}.
\end{abstract}




\section{Introduction}
In a protoplanetary disc, planet formation by core accretion requires the gathering of solid dust into planetesimals.
Of the possible mechanisms, the Streaming Instability \citep[SI,][]{2005ApJ...620..459Y} is an important possible mechanism for gathering dust too large to grow by collisional coagulation into a self-gravitating core \citep{2018haex.bookE.135A}.
The SI grows from the counter-streaming of gas and dust in the disc, but a majority of the studies of this process so far has considered only a single dust size.
For clarity, we will describe this as monodisperse-SI (mSI).
Several groups have run fully nonlinear simulations of SI with multiple particle sizes \citep{2010ApJ...722.1437B,2018A&A...618A..75S}, but recent linear analysis in this regime, examining the limit of a continuum of dust sizes, has produced surprising results. 
\citet{2019ApJ...878L..30K} numerically approached the infinite-fluid limit of a multifluid streaming instability, employing the multifluid formulation of the dust-gas system derived in \citet{2019ApJS..241...25B}.
Surprisingly, it was found that fast growth is limited to narrower regimes of parameter space than in the mSI case \citep{2019ApJ...878L..30K, 2020arXiv200801119Z}.

In the terminology introduced in \citet{paper1} (hereafter \citetalias{paper1}) we denoted the Polydisperse Streaming Instability (PSI) as the generalization of mSI to a dust distribution with a continuum of particle sizes.
Previously, we analyzed the PSI in the short dust stopping time limit with a terminal velocity approximation.
We found that in this limit the PSI takes on a new character and is driven from a resonance which occurs in dust size space, the size resonance.
We demonstrated analytically that for wide, power-law dust size distributions with short stopping times such that the  terminal velocity approximation holds, the PSI grows only at wavenumbers much larger than where the mSI grows fastest. Fast PSI growth (on approximately a dynamical time scale) is restricted to dust to gas mass ratios approximately greater than unity, and we demonstrated the accuracy of these models by comparison to numerical calculations computed with a simple direct discretization of the eigenproblem.

Independently,
\citet{2020arXiv200801119Z}
continued the program of 
\citet{2019ApJS..241...25B} and \citet{2019ApJ...878L..30K} in computing results for multifluid systems approaching the continuous limit.
They also found that fast growth is limited to dust to gas mass ratios approximately greater than unity when exclusively tightly coupled dust is considered, but also found fast growth with top-heavy power-law dust distributions when the peak particle size had an associated stopping time longer than an orbital timescale.

In this paper, we provide a more in-depth derivation of the model of \citetalias{paper1} and present methods for calculating results for the linear PSI stability problem.
The PSI stability problem turns out to be very challenging numerically. 
This was already recognised in \citet{2019ApJ...878L..30K}, where they considered up to $2048$ separate dust fluids, yielding a eigenvalue problem with a matrix of size $8196\times 8196$. Even then, there were cases where they could only find upper limits to the growth rate. Since finding eigenvalues of a dense matrix is an $O(N^3)$ computation (with $N$ the matrix size), increasing the number of fluids becomes computationally prohibitive. In numerical terms and in the context of our model, the approach presented by \citet{2019ApJS..241...25B} yields a first-order discretization of the integrals over the dust distribution which provide the gas-dust coupling in the fluid equations. In \citetalias{paper1} we described a method based on a second-order discretization of these integrals, that has superior convergence behaviour but which still suffers from the same underlying computational limitations.

Because of the computational difficulty, the focus of this paper is entirely on the mathematics of the PSI stability problem and the methods needed to solve it efficiently, leaving discussion of the physics of planet formation and the role of PSI for the subsequent paper in this series.
The formulation and limitations of the governing equations is of central importance to understanding the physical applicability of the linear stability problem, so we present this in detail in this paper.
We then further describe our \citetalias{paper1} method and the difficulties that this family of methods based on direct discretization of the governing system of equations have with the PSI eigenproblem.
With this context, we present an alternate method, which avoids the limitations of the \citet{2019ApJS..241...25B} and \citetalias{paper1} methods, providing far more accurate results at a lower computational cost.

These methods require complex implementations. 
So as a service to the community, we release as a companion to this paper an implementation in the publicly available Python-language {\tt psitools} \citep{psitools} package.
Thus, this paper serves two purposes:
It describes the mathematics of the PSI eigenproblem with  numerical methods required to solve them, and it also serves as high-level documentation of the {\tt psitools} package.
The algorithmic building blocks of the approach used in {\tt psitools} to solve the PSI eigenproblem have all previously existed, and are not special to this problem.
They may find further application in other areas of theoretical astrophysics where stability eigenproblems are not amenable to simple direct discretization techniques.

This paper is organized as follows.
First, in Section~\ref{sec:derivation} we present the derivation for the governing equations for polydisperse dust-gas flows from a Boltzmann equation approach.
In Section~\ref{sec:linear} we present the linear PSI perturbation equations for stability analysis.
Then, in Section~\ref{sec:nummeth} we describe the numerical methods: The properties of the direct eigensolver are discussed in
Section~\ref{sec:direct},
the new root-finding technique is described in 
Section~\ref{sec:rootfinding},
a related root-counting method is described in
Section~\ref{sec:rootcounting},
and 
Section~\ref{sec:mapping} describes an algorithm for efficiently building maps of PSI growth across a continuous parameter space.
In Section~\ref{sec:examples} we give benchmark results for the PSI problem and compare results obtained with these new algorithms to previously published results.
Finally, we summarize the results and give conclusions in Section~\ref{sec:conclusions}.


\section{Derivation of governing equations for polydispserse dust-gas flows}
\label{sec:derivation}

We are interested in the evolution of a mixture of gas and solids (or dust), where the solids have a range of sizes and the number of solid particles is large enough that their size distribution can be thought of as a continuum. The solids and gas interact through a drag force. The gas obeys well-known continuum equations, including mass, momentum and energy conservation, together with equations for magnetic and radiation fields, if present. In order to elucidate the general principles, and also because for the problem at hand it is sufficient, we focus on the simple case of an unmagnetized gas with a barotropic equation of state, which is governed by the usual Euler equations:
\begin{align}
\partial_t\rhog + \nabla\cdot (\rhog \vg) =& 0\, ,\label{eq:gascont}\\
\partial_t\vg + (\vg\cdot\nabla)\vg =& -\frac{\nabla p}{\rhog} +\bm{\alpha}_{\rm g} + \bm{\alpha}_{\mathrm{drag,g}}\, ,
\label{eq:gasmom}
\end{align}
where $\rhog$ is the gas density, $\vg$ the gas velocity, $p$ denotes gas pressure and we have separated the forces acting on the gas into an acceleration due to the drag force coupling gas and dust\footnote{We use $\bm{\alpha}$ for acceleration to avoid confusion with particle size $a$.} $\bm{\alpha}_{\rm drag,g}$, and any other force acting on the gas, leading to an acceleration $\bm{\alpha}_{\rm g}$. For a barotropic fluid, a simple equation of state $p=p(\rhog)$ closes the system, but we note that the formalism below can be trivially extended to more complex equations as long as the only interaction between gas and solids is through the drag force. The only term that depends on the nature of the solid component is $\bm{\alpha}_{\mathrm{drag,g}}$, for which we need to find an expression when the solids have a continuous distribution in size.

Equations (\ref{eq:gascont}) and (\ref{eq:gasmom}) can be derived from a kinetic picture by taking velocity moments of the Boltzmann transport equation \citep[e.g.][]{chapman_cowling_1939} and identifying the pressure as due to a random velocity component. Below, following for example \cite{2004ApJ...603..292G} and \cite{2011MNRAS.415.3591J}, we take the same approach for the solid component of the mixture, which now also has a distribution in particle size.

\subsection{Boltzmann equation}

Consider the distribution function for dust particles $f({\bf x}, {\bf v}, a, t)$ so that
$f({\bf x}, {\bf v}, a, t)\rmd{\bf x} \rmd{\bf v} \rmd a$ is the number of dust particles in a volume $\rmd{\bf x}$ around ${\bf x}$, with velocities in a (velocity) volume $\rmd{\bf v}$ around ${\bf v}$ and with size between $a$ and $a+\rmd a$. Note that the units of $f$ are $[f]=\mathrm{cm}^{-7}~\mathrm{s^3}$. The number of particles per unit volume (i.e. the number density) is given by
\begin{align}
n_{\rm d}({\bf x}, t) = \int\int f({\bf x}, {\bf v}, a, t) \rmd{\bf v} \rmd a,
\end{align}
while the volume density is
\begin{align}
\rhod({\bf x}, t) =\rho_{\rm b} \int\int V(a) f({\bf x}, {\bf v}, a, t) \rmd{\bf v} \rmd a,
\end{align}
where $\rho_{\rm b}$ is the bulk density of a dust particle and $V(a)$ is its volume ($V(a)=4\pi a^3/3$ for spherical dust particles). The evolution of $f$ is given by the nonlinear Boltzmann equation \citep{1872boltzmann}:
\begin{align}
\partial_t f + {\bf v}\cdot \nabla_x f + \nabla_v\cdot (f \bm{\alpha})= 0,
\label{eqBoltzmann}
\end{align}
where the right hand side is zero since we do not consider any collisions between dust particles. Note the subscripts on the gradient operator denoting whether the gradient is to be taken in real space or velocity space. Here $\bm{\alpha}$ is the acceleration on the dust component due to external forces, which can again be split into a component $\bm{\alpha}_{\rm d}$ covering everything but the drag force, and the drag acceleration $\bm{\alpha}_{\rm drag,d}$.

\subsection{Velocity moments}
\label{sec:velmom}

We can take moments of the Boltzmann equation by multiplying by a quantity $\rho_{\rm b} V(a) {\bf v}^p$, where the power index $p$ is to be defined later, and integrating over velocity:
\begin{align}
\int \rho_{\rm b} V(a) {\bf v}^p\left(\partial_t f + {\bf v}\cdot \nabla_x f +
  \nabla_v\cdot(f\bm{\alpha})\right)\rmd{\bf v} =0.
\label{eqBoltz}
\end{align}
The first term can be simplified to
\begin{align}
\int \rho_{\rm b} V(a) {\bf v}^p\partial_t f \rmd{\bf v}  = \partial_t\left(\int \rho_{\rm b} V(a) {\bf v}^p f \rmd{\bf v} \right) \equiv \partial_t(\sigma\left< {\bf v}^p\right>_{\bf v}),
\end{align}
where we define a velocity average of quantity $A$ as 
\begin{align}
\left<A\right>_{\bf v} \equiv \frac{\int Af \rmd{\bf v}}{\int f \rmd{\bf v}},
\end{align}
and where we define the \emph{size density}\footnote{In keeping in line with terms such as `volume mass density', `surface mass density' and `line charge density', $\sigma$ could be called `volume size mass density', but since that in cases where `mass' and `volume' are clear from the context, $\rho$ is known simply as `density', we prefer the shorter term `size density' for $\sigma$.} $\sigma$ as
\begin{align}
\sigma({\bf x}, t, a) = \rho_{\rm b} V(a) \int f({\bf x}, {\bf v}, a, t) \rmd{\bf v}. \label{eq:sizedensity}
\end{align}
The size density is the mass density per unit size, i.e. $\sigma \rmd a$ is the mass density between $a$ and $a +\rmd a$. For a polydisperse fluid, it replaces the mass density in the evolution equations, as we will see below.

In a similar way, the second term in (\ref{eqBoltz}) can be simplified to
\begin{align}
\int \rho_{\rm b} V(a) {\bf v}^p {\bf v}\cdot \nabla_x f \rmd{\bf v}  =
\nabla_x\cdot(\sigma\left< {\bf v}^p{\bf v}\right>_{\bf v}),
\end{align}
and, finally, the third term:
\begin{align}
\int \rho_{\rm b} V(a) {\bf v}^p  \nabla_v\cdot(f\bm{\alpha})\rmd{\bf v}  =\nonumber\\
 \int \left[\nabla_v\cdot(\rho_{\rm b} V(a) {\bf v}^p  \bm{\alpha}f)  - f(\bm{\alpha}\cdot\nabla_v) \rho_{\rm b} V(a) {\bf v}^p\right]\rmd{\bf v}  = \nonumber\\
 -\sigma\left<(\bm{\alpha}\cdot\nabla_v) {\bf v}^p\right>_{\bf v},
 \end{align}
where we have assumed that $f$ goes to zero fast enough as $|{\bf v}|\rightarrow \infty$ so that ${\bf v}^p |\bm{\alpha}| f$ vanishes. We end up with a velocity-integrated version of (\ref{eqBoltz}) that reads:
\begin{align}
\partial_t(\sigma\left< {\bf v}^p\right>_{\bf v}) + \nabla_x\cdot(\sigma \left<{\bf v}^p{\bf v}\right>_{\bf v})- \sigma\left<(\bm{\alpha}\cdot\nabla_v) {\bf v}^p\right>_{\bf v}=0.\label{eqAverage}
\end{align}
The first two moments, $p=0$ and $p=1$ give
\begin{align}
\partial_t\sigma + \nabla_x\cdot(\sigma\left< {\bf v}\right>_{\bf v})=& 0,\label{eqmom1}\\
\partial_t(\sigma\left<{\bf v}\right>_{\bf v}) + \nabla_x\cdot\left(\sigma\left< {\bf v}^2\right>_{\bf v}\right)=& \sigma \left<\bm{\alpha}\right>_{\bf v}.\label{eqmom2}
\end{align}
Rewrite the last equation as
\begin{align}
\partial_t(\sigma\left<{\bf v}\right>_{\bf v})
+& \nabla_x\cdot(\sigma\left< {\bf v}\right>_{\bf v}\left< {\bf v}\right>_{\bf v})
=\nonumber\\
&\sigma \left<\bm{\alpha}\right>_{\bf v}
- \nabla_x\cdot\left(\sigma\left< {\bf v}^2\right>_{\bf v}-\sigma\left< {\bf v}\right>_{\bf v}\left< {\bf v}\right>_{\bf v}
\right),
\label{eq:momstress}
\end{align}
where the last term is the divergence of a stress tensor. The fluid approximation is essentially to neglect this term, which can be done for small enough stopping times \citep{2004ApJ...603..292G, 2011MNRAS.415.3591J}. We provide some further comments on this in appendix \ref{sec:fluid}. Singling out the velocity dependence of $\bm{\alpha}$, if $\bm{\alpha}({\bf v})$ is at most quadratic in ${\bf v}$, it follows immediately that $\left<\bm{\alpha}\right>_{\bf v} = \bm{\alpha}(\left<{\bf v}\right>_{\bf v})$.

If we define the size-dependent bulk velocity $\ud \equiv \left< {\bf v}\right>_{\bf v}$, and drop the subscript $x$ on the gradient operators, we arrive at
\begin{align}
\partial_t\sigma + \nabla\cdot(\sigma \ud )=& 0,\label{eq:dustcont}\\
\partial_t(\sigma \ud)
+ \nabla\cdot(\sigma \ud\ud)
=&
\sigma\bm{\alpha}(\ud)
\label{eq:dustmom}.
\end{align}
Note that these are the ordinary dust fluid equations, with volume density $\rhod$ replaced by the size density $\sigma$ and velocity $\vd$ replaced by a size-dependent velocity $\ud$.

\subsection{Size-integrated equations}

The dust volume density and momentum are given by integrals over dust size:
\begin{align}
\rhod =& \int \sigma \rmd a,\\
\rhod \vd =& \int \sigma \ud \rmd a.
\end{align}
If we integrate the dust fluid equations (\ref{eq:dustcont}) and (\ref{eq:dustmom}) over dust size, we get
\begin{align}
\partial_t\rhod + \nabla\cdot\left(\rhod\vd\right)=& 0,\\
\partial_t\left(\rhod\vd\right) + \nabla\cdot\left(\int\sigma \ud\ud\rmd a\right)=&
 \int\sigma \bm{\alpha}
 \rmd a.
\end{align}
The second equation can be rewritten as
\begin{align}
\partial_t\left(\rhod\vd\right) + 
\nabla\cdot\left(\rhod\vd\vd\right) =  
\int\sigma \bm{\alpha} \rmd a - \nabla\cdot \mathsf{S},
\label{eqDustMomInt}
\end{align}
with stress tensor
\begin{align}
\mathsf{S} = \int\sigma \ud \ud \rmd a-\rhod\vd\vd 
= \int \sigma (\ud - \vd )(\ud+\vd)\rmd a.
\end{align}
Note that this stress tensor measures velocity correlations in size space rather than velocity space, and for wide enough size distributions can not be neglected in general. In \citetalias{paper1} we found that when all particles are well-coupled to the gas, the stress tensor vanishes. 

If we split the acceleration into a drag acceleration $\bm{\alpha}_{\rm drag,d}$ and any other force acting on the dust $\bm{\alpha}_{\rm d}$, we find that the total momentum transfer between gas and dust is
\begin{align}
\int\sigma \bm{\alpha}_{\rm drag,d} \rmd a.
\end{align}
 Conservation of momentum then dictates that the backreaction on the gas in (\ref{eq:gasmom}) must be
\begin{align}
\bm{\alpha}_{\rm drag,g} = -\frac{1}{\rhog}\int \sigma \bm{\alpha}_{\rm drag,d}\rmd a.
\label{eq:gasdrag}
\end{align}
This term allows us to assemble the governing equations. Note that upon taking the size density a delta function in size space, the stress tensor again vanishes and the monodisperse dust-gas equations are recovered.

\subsection{Governing equations}

Armed with the results of the previous subsections, we can now write down the equations governing a mixture of gas and dust particles, where the solids have a continuous size distribution, using equations (\ref{eq:gascont}), (\ref{eq:gasmom}), (\ref{eq:dustcont}), (\ref{eq:dustmom}) and (\ref{eq:gasdrag}):
\begin{align}
\partial_t\rhog + \nabla\cdot (\rhog \vg) =& 0\, ,\label{eq:govfirst}\\
\partial_t\vg + (\vg\cdot\nabla)\vg =& -\frac{\nabla p}{\rhog} +\bm{\alpha}_{\rm g} - \frac{1}{\rhog}\int \sigma \bm{\alpha}_{\rm drag,d}\rmd a\, ,\label{eq:govsecond}
\\
\partial_t\sigma + \nabla\cdot(\sigma \ud )=& 0,\label{eq:govthird}\\
\partial_t\ud
+ (\ud\cdot\nabla)\ud
=&
\bm{\alpha}_{\rm d}  + \bm{\alpha}_{\rm drag,d}.\label{eq:govlast}
\end{align}
The main differences with the two-fluid model are that the drag on the gas is now an integral over size, and that the size density $\sigma$ and the velocity ${\bf u}$ depend on dust size. 
These equations were employed in \citetalias{paper1} and  similar equations were used  in \citet{2005ApJ...625..414T}, though without reference to a derivation.

\subsection{Possible generalizations}

Equations (\ref{eq:govfirst})-(\ref{eq:govlast}), where the equation of state is barotropic, are sufficient to study the streaming instability in its polydisperse form. It should be clear that adding more complicated gas physics is trivial as long as it does not lead to further interactions between gas and solids. For example, one could study the effect of magnetic fields on the gas by adding the Lorentz force to $\bm{\alpha}_{\rm g}$ and including the induction equation. This is a trivial extension unless for example the resistivity depends on dust density, in which case the resistivity will involve an integral over $\sigma$. Another example would be the inclusion of a gas energy equation, which is a trivial extension if gas and dust temperatures can be assumed to be either equal or independent. Again, coupling of dust and gas temperatures will involve an integral over $\sigma$.  

\section{The linear PSI}
\label{sec:linear}

\subsection{Basic equations}

We set up the problem in the well-known geometry of a shearing box \citep{1965MNRAS.130..125G}, which is a local Cartesian coordinate frame, orbiting at radius $r_0$ with angular velocity $\bm{\Omega}$, where the $x$-direction coincides with the radial direction and $y$ with the azimuthal direction, while $z$ is the vertical direction, perpendicular to the disc mid plane, as usual. The body forces acting on gas and dust are the Coriolis force and a tidal force through an effective potential $\Phi=-S\Omega x^2$, where $S$ is the shear rate of the disc ($S=3\Omega/2$ in a Keplerian disc). In addition, we give the gas an extra acceleration in the $x$ direction in order to accommodate effects from a global pressure gradient \citep{2005ApJ...620..459Y,2007ApJ...662..613Y}, so that the total acceleration of the gas is given by
\begin{align}
\bm{\alpha}_{\rm g} = 2\eta {\bf \hat x} - 2\bm{\Omega}\times {\bf v}_g-\nabla\Phi.
\end{align} 
Since the parameter governing the sub-Keplerian nature of the disc comes from a \emph{global} pressure gradient, for the local model $\eta$ has to be prescribed as an input parameter. If we denote the global pressure by $P$, we have that
\begin{align}
\eta = \frac{1}{2\rhog}\frac{\partial P}{\partial r} \sim \frac{c^2}{r_0},
\end{align} 
where $c$ is the speed of sound in the gas and $r_0$ is the fiducial orbital radius of the shearing box. The acceleration on the dust is the same but with the term involving $\eta$ omitted. 

We take the drag force to be in the Epstein regime:
\begin{align}
\bm{\alpha}_{\rm drag,d} = - \frac{\ud-\vg}{\taus(a)},
\end{align}
where $\taus$ is the (size-dependent) \emph{particle} stopping time \citep[see e.g.][]{2003A&A...399..297W} and therefore depends on gas density:
\begin{align}
\taus = \sqrt{\frac{\pi}{8}}\frac{a\rho_{\rm b}}{\rhog c}.
\end{align}
We note that the gas density dependence is sometimes neglected in calculations of the streaming instability, either explicitly \citep{2019ApJ...878L..30K} or through an incompressibility condition on the gas \citep{2005ApJ...620..459Y}. Since the gas motions involved are very subsonic, this should lead to minor differences only \citep{2007ApJ...662..613Y}. Using a reference gas density, one can define a Stokes number\footnote{Different authors have used different notations for stopping times and Stokes numbers. To avoid possible confusion, we note that while \cite{2017ApJ...849..129L} use our definition of $\taus$, \cite{2005ApJ...620..459Y}, \cite{2018MNRAS.477.5011S} and \cite{2019arXiv190605371U} use $t_{\rm stop}$ for the particle stopping time and $\taus$ for the Stokes number. Other conventions are by \cite{2019ApJ...878L..30K}, who use $t_s$ for particle stopping time and $T_s$ for Stokes number, and by \cite{2020MNRAS.492.4591J}, who use $t_{\rm stop}$ for particle stopping time and $\mathrm{St}$ for a (modified) Stokes number.} through
\begin{align}
\mathrm{St}=\Omega \taus.
\end{align}
Tightly coupled particles have $\mathrm{St} \ll 1$, while the strongest drift velocities occur for particles with $\mathrm{St} \sim 1$ \citep[e.g.][]{1977MNRAS.180...57W}. 

The equations governing the gas are then given by (\ref{eq:govfirst}) and (\ref{eq:govsecond}), with accelerations as stated above:
\begin{align}
\partial_t\rhog + \nabla\cdot(\rhog\vg) =& 0\, ,\label{eq:gascont_sbox}\\
\partial_t \vg
+ (\vg \cdot \nabla)\vg
=&
2\eta {\bf \hat x} -\frac{\nabla p}{\rhog}- 2\bm{\Omega}\times \vg-\nabla\Phi \nonumber\\
&+ \frac{1}{\rhog}\int \sigma\frac{\ud-\vg}{\taus(a)}\rmd a.\label{eq:gasmom_sbox}
\end{align}
We take the equation of state to be isothermal, $p = c^2\rhog$, with constant sound speed $c$. The dust fluid equations can similarly be constructed from (\ref{eq:govthird}) and (\ref{eq:govlast}):
\begin{align}
\partial_t\sigma + \nabla\cdot(\sigma \ud )=& 0,\label{eq:dustcont_sbox}\\
\partial_t \ud
+ (\ud \cdot \nabla)\ud
=&
-2\bm{\Omega}\times \ud-\nabla\Phi - \frac{\ud-\vg}{\taus(a)}.\label{eq:dustmom_sbox}
\end{align}
With these shearing-box equations for the gas-dust system in hand, we can proceed to solve for the equilibrium state.

\subsection{Equilibrium state}

For the equilibrium state, we take $\rhog$ and $\sigma$ to be spatially constant. An equilibrium solution can then be found with velocities independent of $y$ and $z$ and no vertical velocity (\citetalias{paper1}; see also \citealt{2005ApJ...625..414T}, \citealt{2018MNRAS.479.4187D}): 
\begin{align}
\vgx
=&
\frac{2\eta}{\kappa}\frac{\mathcal{J}_1}{\left(1+ \mathcal{J}_0\right)^2 + \mathcal{J}_1^2},
\\
v_{gy}
=&
-Sx -\frac{\eta}{\Omega}\frac{1 +  \mathcal{J}_0}{\left(1+ \mathcal{J}_0\right)^2 + \mathcal{J}_1^2},\\
u_x =& \frac{2\eta}{\kappa}
 \frac{\mathcal{J}_1 - \kappa\taus(a)(1 +  \mathcal{J}_0)}{(1+\kappa^2\taus(a)^2)(\left(1+ \mathcal{J}_0\right)^2 + \mathcal{J}_1^2)} ,\\
 u_y   =& -Sx
- \frac{\eta}{\Omega}\frac{1 +  \mathcal{J}_0 + \kappa\taus(a)\mathcal{J}_1}{(1+\kappa^2\taus(a)^2)(\left(1+ \mathcal{J}_0\right)^2 + \mathcal{J}_1^2)},
\end{align}
with integrals
\begin{align}
\mathcal{J}_m = \frac{1}{\rhog}\int \frac{\sigma (\kappa\taus(a))^{m}}{1+\kappa^2\taus(a)^2} \rmd a.
\end{align}
Here $\kappa$ is the epicyclic frequency. For $\eta=0$, only the linear shear remains for both gas and dust, while $\eta\neq 0$ introduces drift between gas and dust. It is worth noting that for tightly coupled particles, the radial drift velocity is first order in $\mathrm{St}$, while the azimuthal drift velocity is second order. Also, note that as mentioned in section \ref{sec:fluid} this equilibrium state is in principle valid outside the regime of validity of the fluid approximation, as it has no velocity dispersion. However, it should be kept in mind that this is probably best viewed as an artefact of the unstratified, unmagnetised shearing box.  

\subsection{Linear perturbations}

Consider small Eulerian perturbations on top of the equilibrium state, $X({\bf x}, t, a) = X^0({\bf x}, a) + X^1({\bf x}, t, a)$, where $X$ stands for any hydrodynamic variable (noting that gas quantities do not depend on dust size), a superscript '0' indicates the equilibrium state and a superscript '1' indicates a small perturbation. Keeping terms up to linear order in perturbed quantities, and noting that $\rhog\taus$ is constant for Epstein drag, we find from (\ref{eq:gascont_sbox}) and (\ref{eq:gasmom_sbox}) that the gas perturbations are governed by:
\begin{align}
\partial_t\rhogo
&+ \vge\cdot \nabla\rhogo
+ \rhogn\nabla\cdot \vgo = 0,\\
\partial_t \vgo
&+ \vgxn \partial_x\vgo
-S \vgxo{\bf \hat y}
=
-\frac{\nabla p^1}{\rhogn}- 2\bm{\Omega}\times \vgo\nonumber\\
&+ \frac{1}{\rhogn}\int \sigma^1\frac{\Delta\ud^0}{\taus(a)}\rmd a
+ \frac{1}{\rhogn}\int \sigma^0\frac{\ud^1-\vgo}{\taus(a)}\rmd a,
\end{align}
with equilibrium relative velocity $\Delta \ud^0 = \ud^0-\vge$. From (\ref{eq:dustcont_sbox}) and (\ref{eq:dustmom_sbox}) we find that dust perturbations are governed by:
\begin{align}
\partial_t\sigma^1
&+ \ud^0\cdot\nabla\sigma^1
+ \sigma^0\nabla\cdot \ud^1 = 0,\\
\partial_t \ud^1
&+ u_x^0 \partial_x\ud^1
 - Su_{x}^1{\bf \hat y}
=\nonumber\\
&-2\bm{\Omega}\times \ud^1
- \frac{\ud^1-\vge}{\taus(a)}
- \frac{\rhogo}{\rhogn} \frac{\Delta \ud^0}{\taus(a)}.
\end{align}
We consider perturbations of the form $X^1({\bf x},t,a) = \hat X(a) \exp(\rmi {\bf k}\cdot {\bf x} - \rmi \omega t)$, with wavenumber ${\bf k} = (k_x, k_y, k_z)^T$ and frequency $\omega$. Note that if $\omega$ has a positive imaginary part this signals exponential growth\footnote{This follows the convention of \cite{2005ApJ...620..459Y}, who define a growth rate $s=\Im(\omega)$, \cite{2017ApJ...849..129L}, whose $\sigma$ corresponds to our $\omega$, \cite{2018MNRAS.477.5011S}, and \cite{2020MNRAS.492.4591J}. We caution the reader that \cite{2019ApJ...878L..30K} consider perturbations with a different time dependence, $\exp(-\omega t)$, as well as \cite{2020ApJ...891..132C}, who use $\exp(\sigma t)$.}. We consider only perturbations with $k_y=0$ (axisymmetric in a global context). Using the above form of the perturbations, the dust and gas perturbation equations transform to:
\begin{align}
k_x \vgxn \frac{\hatrhog}{\rhogn}
&+ {\bf k}\cdot \hatvg =
\omega \frac{\hatrhog}{\rhogn},\label{eq:eig_gasdens}\\
k_x \vgxn \hatvg
&+ \rmi S \hatvgx{\bf \hat y}
+\frac{{\bf k} c^2\hatrhog}{\rhogn}
- 2\rmi \bm{\Omega}\times \hatvg\nonumber\\
&+ \frac{\rmi}{\rhogn}\int \hat \sigma\frac{\Delta{\bf u}^0}{\taus(a)}\rmd a
+ \frac{\rmi}{\rhogn}\int \sigma^0\frac{{\bf \hat u}-\hatvg}{\taus(a)}\rmd a
=
\omega \hatvg,\label{eq:eig_gasvel}\\
 k_xu_x^0\hat \sigma
&+ \sigma^0{\bf k}\cdot {\bf \hat u} =
\omega \hat\sigma,\label{eq:eig_dustdens}\\
k_x u_x^0 {\bf \hat u}
&+ \rmi S\hat u_{x}{\bf \hat y}
-2\rmi\bm{\Omega}\times {\bf \hat u}
-\rmi \frac{{\bf \hat u}-\hatvg}{\taus(a)}
-\rmi \frac{\hatrhog}{\rhogn} \frac{\Delta {\bf u}^0}{\taus(a)}.
=
\omega {\bf \hat u}.\label{eq:eig_dustvel}
\end{align}
These equations constitute an integral eigenvalue problem for eigenvalue $\omega$. It is often convenient to work with dimensionless units by choosing a time scale $\Omega^{-1}$ and a length scale $\eta/\Omega^2$. The parameters governing the system are then the non-dimensional wave vector ${\bf K} = {\bf k}\eta/\Omega^2$, the shear parameter $S/\Omega$, the dust to gas ratio $\mu = \rhodn/\rhogn$, the non-dimensional gas sound speed $c/(\Omega\eta)$ and the size density $\sigma^0(a)$. 

It should be noted that both size density $\sigma$ and dust velocity ${\bf u}$ depend on dust size, background as well as perturbations. By taking the size density a delta function in size space, we recover the monodisperse eigenvalue problem for the streaming instability after multiplying (\ref{eq:eig_dustvel}) by $\sigma^0$ and integrating the dust equations over size. 

Similar to its monodisperse counterpart, PSI modes are largely incompressible. Taking the incompressible limit of (\ref{eq:eig_gasdens})-(\ref{eq:eig_dustvel}) leads to a simplified problem in the terminal velocity approximation \citepalias{paper1}. It does not considerably simplify the full problem, however, as the main numerical difficulties arise in accurately evaluating the integrals over size.

\subsection{Size resonance}
A important result from analysing the PSI linear stability problem in the terminal velocity limit obtained in \citetalias{paper1} was the existence and role of the size resonance. 
This can be simply expressed as the radial phase velocity of the mode matching the background radial drift velocity of the dust. More precisely, in dust stopping time ($\taus$) space it is located where
\begin{align}
    \frac{\Re(\omega)}{k_x} = \uxn(\taus)
    \label{eq:sizeresonance}
\end{align}
is satisfied. From equation~(\ref{eq:eig_dustdens}), at this size we can expect a strong response in the size density. 

The size resonance is closely related to a resonance that comes up in the theory of Resonant Drag Instabilities \citep[RDIs,][]{2018MNRAS.477.5011S,2018ApJ...856L..15S}. RDI theory works for a monodisperse dust fluid in the limit $\mu \ll 1$, so that the backreaction on the gas can be treated as a perturbation. If a wave in the gas can be identified whose phase speed for a particular wavenumber matches the dust advection speed, one expects the strongest reaction, or fastest growth rates for an instability, at precisely this wavenumber. At this resonance it is possible to calculate approximate growth rates in a particular simple way \citep{2018ApJ...856L..15S}. The size resonance, on the other hand, plays a role for \emph{all} values of $\mu$ \citepalias{paper1}. The resonance condition is usually only satisfied by a single size, and even if the resonant size is actually in the size distribution under consideration, the size resonance does not always promote instability \citepalias{paper1}. Unlike the resonance in RDI theory, the size resonance therefore has limited predictive power.

\section{Numerical Methods}
\label{sec:nummeth}

This section presents the numerical methods use to solve the linear PSI stability problem. The PSI eigenvalue problem proves to have significant difficulty. We employ two approaches for solving the PSI linear stability eigenproblem, each with different strengths, to build confidence in the results found.

First, we employ the same direct solver (section \ref{sec:direct}), based on discretizing the dust eigenfunctions as a function of stopping time as in \citetalias{paper1}.
This provides well determined results, but has the disadvantage of very large
computational cost and slow convergence in cases when the instability growth is very small.

Second, we employ a root finding technique to find the roots of the dispersion relation resulting from Equations~(\ref{eq:eig_gasdens}--\ref{eq:eig_dustvel}) directly without discretization (section \ref{sec:rootfinding}). To do this requires high quality estimates for the starting point of the complex root finding iteration. The process of deriving good guesses becomes inherently stochastic, and while when a root is found it is verifiably real and highly accurate, it is possible to miss roots.
Like with the direct solver, this difficulty arises primarily when the instability growth rate is very small, and we will explore the structure of the dispersion relation in the complex plane which leads to this difficulty in detail.

In Section~\ref{sec:rootcounting} we describe the application of a method for counting the roots of the inside a contour in the complex plane.
The application of this method is to provide a separate check on the existence of a growing mode when mapping wavenumber space.
We additionally describe an efficient algorithm for generating maps in wavenumber space of the fastest growing PSI mode, with the root finding algorithm in Section~\ref{sec:mapping}. In all examples, we focus on the traditional MRN dust size distribution \citep{1977ApJ...217..425M}, which was also considered in previous works (\citetalias{2019ApJ...878L..30K, paper1}; \citealt{2020arXiv200801119Z}).

\subsection{Direct Solver}
\label{sec:direct}

\begin{figure}
	\includegraphics[width=\columnwidth]{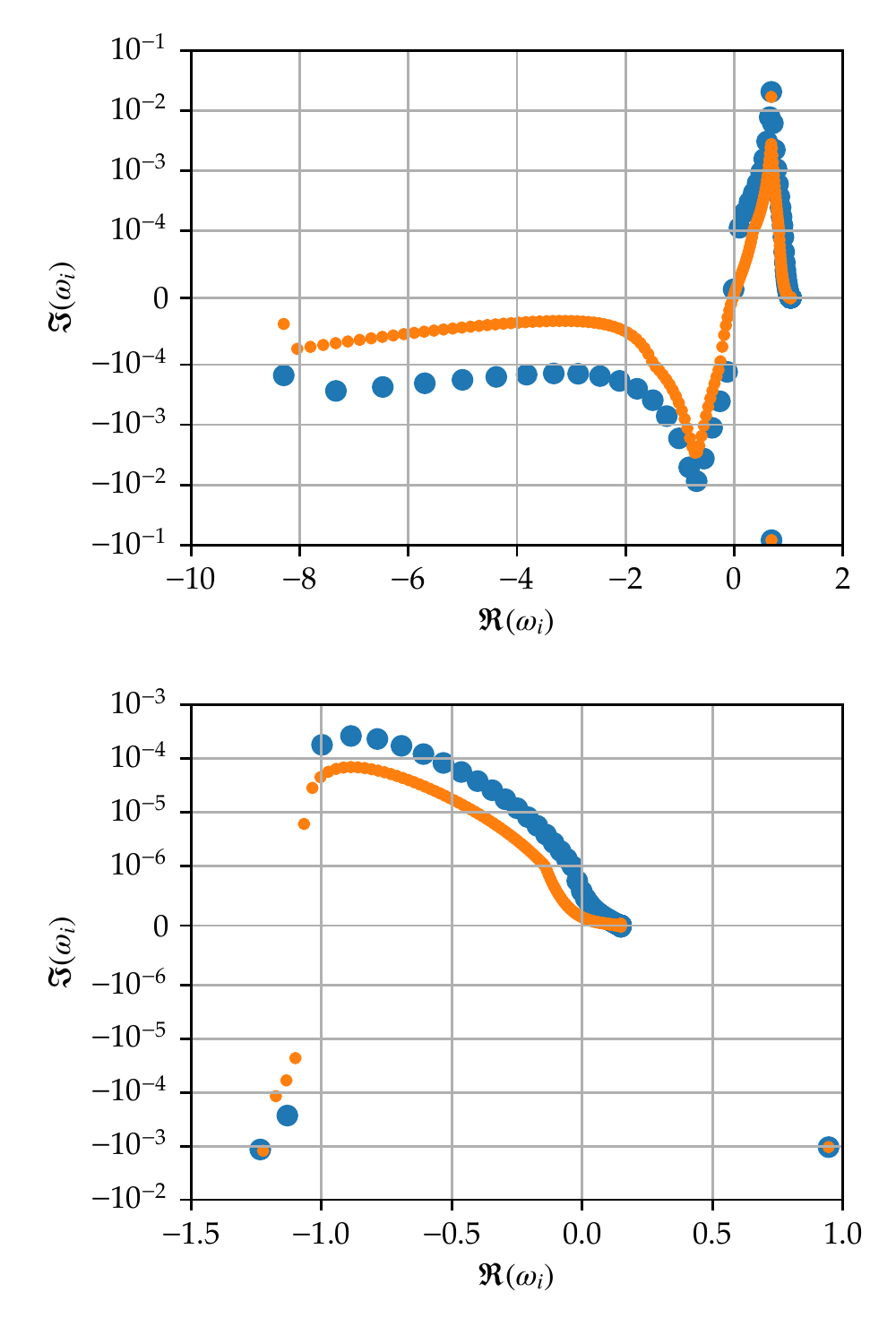}
    \caption{Central section of the distribution of direct solver eigenvalues ($\omega_i$) for {\sl Upper:} $\taus=[10^{-8}, 10^{-2}]\Omega^{-1}$, $\mu=0.5$, $K_x=700$, $K_z=1000$, where the uppermost eigenvalue corresponds to a real physical growing mode and {\sl Lower:} $K_x=100$, $K_z=1000$ where the upper branch does not appear to contain any physical growing modes, both with MRN distributions.
    {\sl Orange:} $L=128$ points {\sl Blue:} $L=512$ points, both with logarithmic gridding in $\taus$. }
    \label{fig:direct_eigenspace}
\end{figure}

\begin{figure}
	\includegraphics[width=0.9\columnwidth]{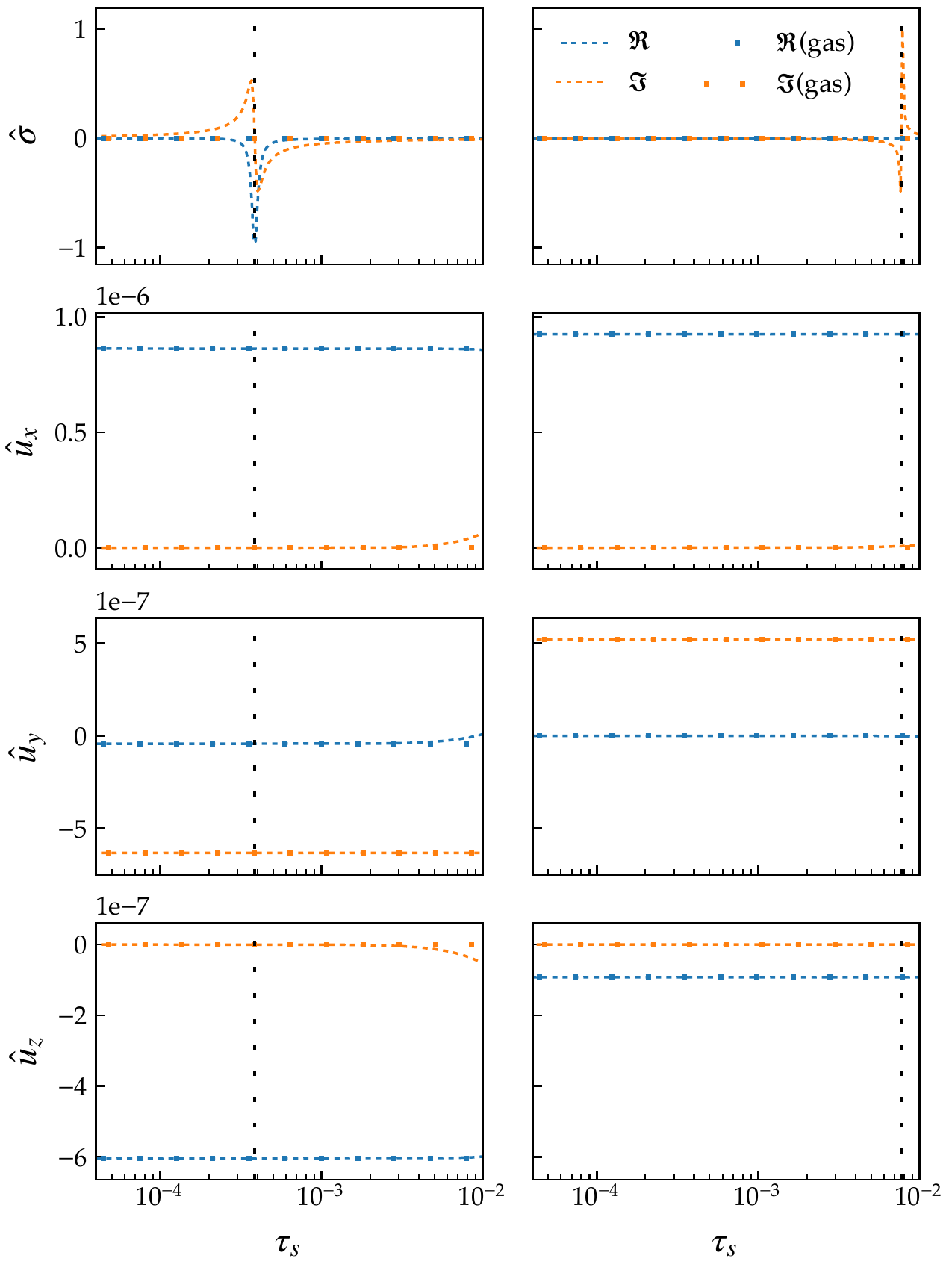}
    \caption{Fastest growing numerical dust eigenfunctions, showing only $\taus>4\times10^{-5}$ from direct solver at $L=512$ on a logarithmic grid. {\sl Left:} $\taus=[10^{-8}, 10^{-2}]\Omega^{-1}$, $\mu=0.5$, $K_x=700$, $K_z=1000$, and {\sl Right:} $K_x=100$, $K_z=1000$ with MRN distributions.
    {\sl Black Dashed}: Position of terminal velocity size resonance. Eigenfunctions are normalized to set $\Im(\hatvgx)=0$ and $|\max(\hatsigma)|=1$.}
    \label{fig:direct_eigenfunctions}
\end{figure}

\begin{figure}
	\includegraphics[width=0.9\columnwidth]{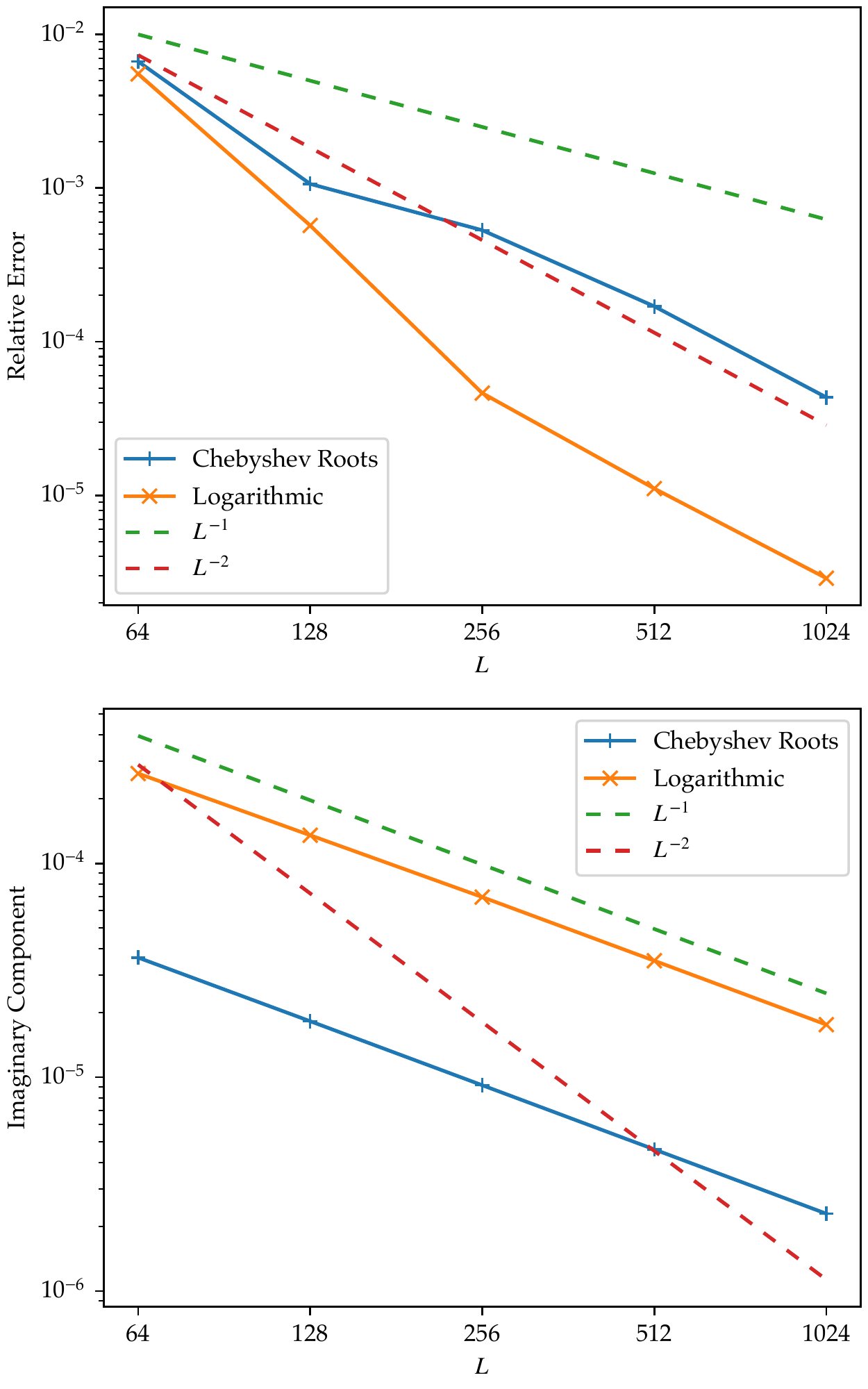}
    \caption{
    Convergence of the direct solver, for smooth and non-smooth eigenfunctions.
    {\sl Upper:} $\taus=[10^{-8}, 10^{-1}]\Omega^{-1}$, $\mu=3$, $K_x=70$, $K_z=100$, a case where the fastest growing eigenmode is smooth and has well resolved finite growth. {\sl Lower:} 
    $K_x=1000$, $K_z=100$ where the numerical fastest growing mode has a sharp discontinuity which leads to first order convergence, apparently towards a zero imaginary part.}
    \label{fig:direct_convergence}
\end{figure}

The most straightforward way to solve the eigenvalue problem is through directly transforming equations~(\ref{eq:eig_gasdens})--(\ref{eq:eig_dustvel}) into a matrix form, and them numerically solving for eigenvalues and eigenvectors.
Our direct solver accomplished this by discretization in $\taus$ and replacing the integrals with a discrete quadrature approximation.
After discretizing on $L$ points in $\taus$ space the results are $4+4L$ eigenvalues $\omega_{L,0\dots 4+4L}$
each with an eigenvector consisting of gas components $\hatrhog$, ${\bf \hat{v}}_{\rm g}$ and the dust eigenfunctions
$\hatsigma(\taus)$ and ${\bf \hat{u}}(\taus)$. 
An eigenvector decomposition of this complex-valued matrix is then found with the {\tt scipy.linalg.eig} routine.
Our implementation  was previously described in \citetalias{paper1}.
The {\tt psitools.direct} module provides this solver with trapezoidal rule quadrature and a small variety of grid functions. Here we describe in more detail the performance of this solver and where the main difficulties with this direct approach to calculating eigenvalues lies.

The direct solver generates 4 additional eigenvalues per added $\taus$-space point used in the discretization, and these do not need to all correspond to physical eigenvalues of the continuous problem \citep{Boyd}.
In Figure~\ref{fig:direct_eigenspace} we show the central section of eigenspace for two different modes, with the numerical eigenvalues for $L=128$ and $L=512$.
In the upper panel a PSI mode with strong growth exists, and the corresponding eigenvalue, with the largest imaginary part, is insensitive to the direct solver resolution.
However, in the lower panel, a mode with no fast PSI growth was chosen, so the fastest growing eigenvalues at each resolution are very sensitive to resolution and the imaginary parts (growth rate) decrease rapidly as resolution is increased. These fastest growing eigenvalues are numerically spurious, and arise from the discretization, not the physical system.

Thus the difficulty with the direct solver when applied to PSI modes with little physical growth lies in either identifying and removing the numerically spurious eigenvalues, or in keeping them well enough resolved to lie below the physical ones in the complex plane.
Reducing the growth rate of the spurious eigenmodes is in turn difficult due to the structure of the associated eigenfunctions.

Shown in Figure~\ref{fig:direct_eigenfunctions} are the dust eigenfunctions for the fastest growing modes in the two cases from Figure~\ref{fig:direct_eigenspace}.
In the formulation of the direct eigensolver, these functions appear in the integrals which are approximated with a trapezoid rule quadrature.
The resolution requirement for the direct solver is apparent from the sharp shape of the $\sigma(\taus)$ functions.
Overplotted in Figure~\ref{fig:direct_eigenfunctions} is the location of the size resonance as determined from the terminal velocity limit.
This can be expressed as the phase speed of the unstable mode matching the radial drift velocity of the dust (see (\ref{eq:sizeresonance})). In both cases, the sharp feature in the dust eigenfunction coincides with the size resonance.
However, on the left the physical growing mode contains a finite-width feature in $\hat{\sigma}$ which becomes better resolved with increasing $\taus$ points, while on the right the feature narrows with higher resolution, and the growth rate of the mode decreases.
In this numerically spurious case, the growth at the size resonance is an artifact of the finite resolution in $\taus$.

On the left of Figure~\ref{fig:direct_eigenfunctions} , the sharp feature in $\sigma$ is physical, and can be reasonably resolved, so the trapezoid rule quadrature converges at second order.
However, on the right, the eigenfunction for the numerically spurious eigenvalue contains an asymptote which narrows continuously as the resolution in $\taus$ is increased, and thus the convergence of the trapezoid rule quadrature is reduced to first order.
This is demonstrated in Figure~\ref{fig:direct_convergence}, where the convergence behaviour of the direct solver is shown for these two cases.
The fastest growing mode converges at second order for the first case, and at first order (towards zero) in the second case where the fastest growing mode is numerically spurious.
Figure~\ref{fig:direct_convergence} also illustrates the impact on the direct solver of the choice of $\taus$ gridding.
In the case of a physically growing eigenvalue, the error obtained with a logarithmic grid in $\taus$ is roughly an order of magnitude lower than with the Chebyshev roots grid.
However, in the case of a numerically spurious eigenvalue, the Chebyshev roots grid outperforms the lograrithmic grid by nearly an order of magnitude.

Finally, there are two practical limits on the accuracy of the direct solver. As it uses a dense matrix eigenvalue solver, the computational cost scales as $O(L^3)$, which with contemporary processors makes $L\sim 4096$ a practical upper limit to the resolution. 
We also find that the roundoff error limitations of double precision floating point begin to infect results for small growth rates at resolutions of this magnitude, so larger resolutions (and smaller errors) cannot be obtained unless higher precision floating point techniques are implemented in the matrix eigensolver, further increasing the computational cost.

The difficulties with the direct solver, inherent to its internal construction of eigenfunctions simultaneously while calculating the eigenvalues we seek, motivates us to formulate a entirely different solution method for the eigenvalue problem, based on directly finding roots of the PSI dispersion relation in the complex plane. 
This side-steps the difficulties of the direct solver, and also provides the ability to cross-check the results obtained with either method.
However, the direct solver converges quickly at second order in cases with fast PSI growth, and the smoothness of the numerically determined growth rate as as a function of wavenumber or other underlying parameters still make it useful when optimizing for maximal growth rates. The ability to construct eigenfunctions is also useful for examining the nature of specific growing modes.

\subsection{Root finding algorithm}
\label{sec:rootfinding}

Motivated by the difficulties of the direct solver approach, we have developed a  root finding algorithm 
for the complex frequency $\omega$  in 
equations~(\ref{eq:eig_gasdens})-(\ref{eq:eig_dustvel}).
The interface to this is implemented in the {\tt psitools.psimode} module. 
This development begins with first casting the linear problem into a single equation, that is, the dispersion relation.

\subsubsection{Dispersion relation in matrix form}

We can obtain an expression for the dispersion relation by eliminating all quantities except $\hatvg$. Starting from equations (\ref{eq:eig_gasdens})-(\ref{eq:eig_dustvel}), first eliminate the gas density perturbation through (\ref{eq:eig_gasdens}):
\begin{align}
\frac{\hatrhog}{\rhogn} = \frac{{\bf k}\cdot \hatvg}{\omega - k_x \vgxn},
\end{align}
and write the gas momentum equation as
\begin{eqnarray}
\mathsf{P} \hatvg
+ \frac{\rmi}{\rhogn}\int \hat \sigma\frac{\Delta{\bf u}^0}{\taus(a)}\rmd a
+ \frac{\rmi}{\rhogn}\int \sigma^0\frac{{\bf \hat u}-\hatvg}{\taus(a)}\rmd a
=0\label{eq:gasmomP},
\end{eqnarray}
where the matrix $\mathsf{P}$ is given by
\begin{eqnarray}
\mathsf{P} = \left(\begin{array}{ccc}
- \omega_{\rm g}  + \frac{k_x^2c^2}{\omega_{\rm g}} & 2\rmi\Omega &  \frac{k_xk_zc^2}{\omega_{\rm g}} \\
-\rmi (2\Omega - S) & - \omega_{\rm g} & 0\\
\frac{k_xk_zc^2}{\omega_{\rm g}}  & 0 & - \omega_{\rm g} + \frac{k_z^2c^2}{\omega_{\rm g}}
\end{array}\right),
\end{eqnarray}
with shifted frequency $\omega_{\rm g} = \omega -k_x\vgxn$. The gas drag terms in the gas momentum equation (\ref{eq:gasmomP}) read:
\begin{align}
\frac{\rmi}{\rhogn}\int \frac{\sigma^0}{\taus(a)} & \left[\Delta{\bf u}^0 \frac{\hat\sigma}{\sigma^0} + {\bf \hat u} - \hatvg\right]\rmd a
=\nonumber\\
& \int \mathcal{K}(a)\left[\Delta{\bf u}^0 \frac{{\bf k}\cdot {\bf\hat u}}{\omega-k_xu_x^0(a)} + {\bf \hat u} - \hatvg\right]\rmd a,
\label{eq:gasdragint}
\end{align}
with kernel $\mathcal{K}= \rmi \sigma^0/(\rhogn\taus)$ and we have used the dust continuity equation to write $\hat\sigma$ in terms of ${\bf \hat u}$. If we define a matrix $\mathsf{V}$ such that
\begin{align}
\int \mathcal{K}(a)\left[\Delta{\bf u}^0 \frac{\hat\sigma}{\sigma^0} + {\bf \hat u} - \hatvg\right]\rmd a
=
\int \mathcal{K}(a)\left[\mathsf{V}(a){\bf \hat u} - \hatvg\right]\rmd a.\label{eq:gasmomV}
\end{align}
It is easily verified that we need
\begin{align}
\mathsf{V} = \mathsf{I} + \frac{1}{\omega-k_xu_x^0}\left(\begin{array}{ccc}
\Delta u_x^0 k_x & 0 & \Delta u_x^0 k_z\\
\Delta u_y^0 k_x & 0 & \Delta u_y^0 k_z\\
0 & 0 & 0
\end{array}\right).
\end{align}
We want to get an expression for ${\bf \hat u}$ in terms of $\hatvg$. The dust momentum equation (\ref{eq:eig_dustvel}) gives, after eliminating gas density:
\begin{align}
\left(k_x u_x^0-\omega-\frac{\rmi}{\taus(a)}\right) {\bf \hat u}
+ \rmi S\hat u_{x}{\bf \hat y}
&-2\rmi\bm{\Omega}\times {\bf \hat u}
=\nonumber\\
&\rmi \frac{{\bf k}\cdot \hatvg}{\omega_{\rm g}}\frac{\Delta {\bf u}^0}{\taus(a)}
-\rmi \frac{\hatvg}{\tau_s(a)}.
\end{align}
Write as matrix equation
\begin{align}
\mathsf{A}(a){\bf\hat u} = \mathsf{D}(a) \hatvg,
\end{align}
with
\begin{align}
\mathsf{A}=\left(\begin{array}{ccc}
d & 2\rmi\Omega & 0\\
\rmi(S-2\Omega) & d & 0\\
0 & 0 & d
\end{array}\right),
\end{align}
with $d = k_xu_x^0 - \omega - \rmi/\taus$, and
\begin{align}
\mathsf{D} = -\frac{\rmi}{\taus}\mathsf{I} + \frac{\rmi}{\taus\omega_{\rm g}}\left(\begin{array}{ccc}
\Delta u_x^0 k_x & 0 & \Delta u_x^0 k_z\\
\Delta u_y^0 k_x & 0 & \Delta u_y^0 k_z\\
0 & 0 & 0
\end{array}\right).
\end{align}
Hence ${\bf \hat u} = \mathsf{A}^{-1}\mathsf{D} \hatvg$, which we can use in (\ref{eq:gasmomV}) to obtain
\begin{align}
\int \mathcal{K}(a) & \left[\Delta{\bf u}^0 \frac{\hat\sigma}{\sigma^0} + {\bf \hat u} - \hatvg\right]\rmd a
=\nonumber\\
& \int \mathcal{K}(a)\left[\mathsf{V}(a)\mathsf{A}^{-1}(a)\mathsf{D}(a) - \mathsf{I}\right]\rmd a  \hatvg \equiv \mathsf{M} \hatvg.
\label{eq:gasmomM}
\end{align}
The inverse of $\mathsf{A}$ is straightforward to calculate:
\begin{eqnarray}
\mathsf{A}^{-1}=\left(\begin{array}{ccc}
-\frac{d}{\kappa^2-d^2} & \frac{2\rmi\Omega}{\kappa^2-d^2} & 0\\
\frac{\rmi(S-2\Omega)}{\kappa^2-d^2} & -\frac{d}{\kappa^2-d^2} & 0\\
0 & 0 & \frac{1}{d}
\end{array}\right).
\end{eqnarray}
The dispersion relation is found by plugging (\ref{eq:gasmomM}) into (\ref{eq:gasmomP}) and is given by 
\begin{align}
\det(\mathsf{P} + \mathsf{M})\equiv f_{\rm disp}(\omega, k_x, k_z, \mu, c, \eta, \Omega, S, \sigma^0(a)) =0,
\end{align}
where we have explicitly indicated the dependence of $f_{\rm disp}$ on the parameters of the system.

It is often convenient to work in non-dimensional units, choosing a time scale $\Omega^{-1}$ and a length scale $\eta/\Omega^2$. The non-dimensional dispersion relation is then
\begin{align}
f_{\rm disp}\left(\frac{\omega}{\Omega}, \frac{\eta k_x}{\Omega^2}, \frac{\eta k_z}{\Omega^2}, \mu, \frac{\Omega c}{\eta}, \frac{S}{\Omega}, \sigma^0(a)\right) =0,
\end{align}
with the understanding that $\sigma^0$ is normalized in such a way that when integrated over size it gives the required dust to gas ratio $\mu$. If we furthermore restrict ourselves to Keplerian discs ($S=3\Omega/2$) and the standard value $\Omega c/\eta = 1/0.05$ \citep[e.g.][]{2005ApJ...620..459Y}\footnote{The specific choice of $\Omega c/\eta$ is not important as long as $\Omega c/ \eta  \gg 1$, so that gas motions are very subsonic and the gas can in principle be treated as an incompressible fluid \citep{2007ApJ...662..613Y}.}, it is clear that the parameters governing the system are $\mu$ and the size distribution $\sigma^0$:
\begin{align}
f_{\rm disp}\left(\nu, K_x, K_z, \mu, \sigma^0(a)\right) =0,
\end{align}
with $\nu = \omega/\Omega$ and $K = \eta k/\Omega^2$.  

\subsubsection{Evaluating the dispersion relation}

Evaluating $f_{\rm disp}$ is an expensive operation because of the integrals in $\mathsf{M}$. Moreover, it turns out the integrals involved can often be very slowly converging, which makes standard numerical quadratures fail. As can be seen from (\ref{eq:gasdragint}), some of the integrands have $\omega - k_x u_x^0(a)$ in the denominator. If $\Re(\omega)=k_xu_x^0(a)$ for some $a$ in the size range under consideration, then the integral diverges for  $\Im(\omega)=0$, and converges very slowly when $\omega$ has a very small imaginary part\footnote{Note that we evaluate the dispersion relation at a particular value of $\omega$, which is therefore a known quantity. What is not known is whether this particular value is in fact an eigenvalue.} . 

\begin{figure}
	\includegraphics[width=\columnwidth]{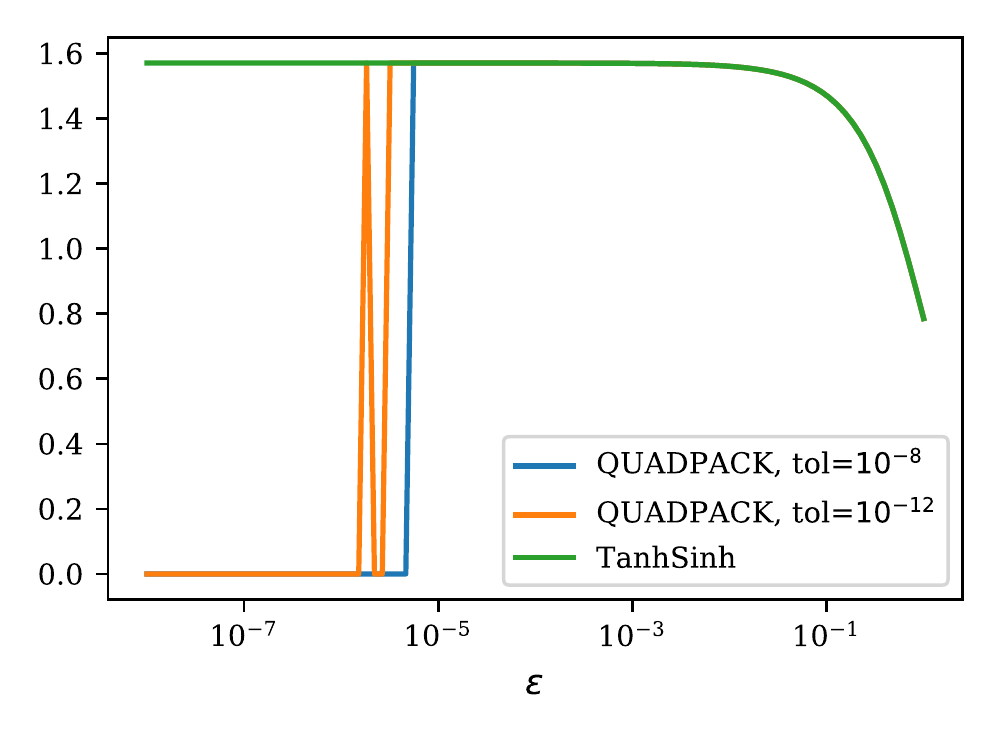}
    \caption{Evaluating the integral (\ref{eq:quadrature}) with \quadpack{} as implemented in {\tt scipy.integrate.quad} at two different error tolerances, together with the result obtained using the tanhsinh quadrature.}
    \label{fig:quadrature}
\end{figure}

\begin{figure*}
	\includegraphics[width=0.9\textwidth]{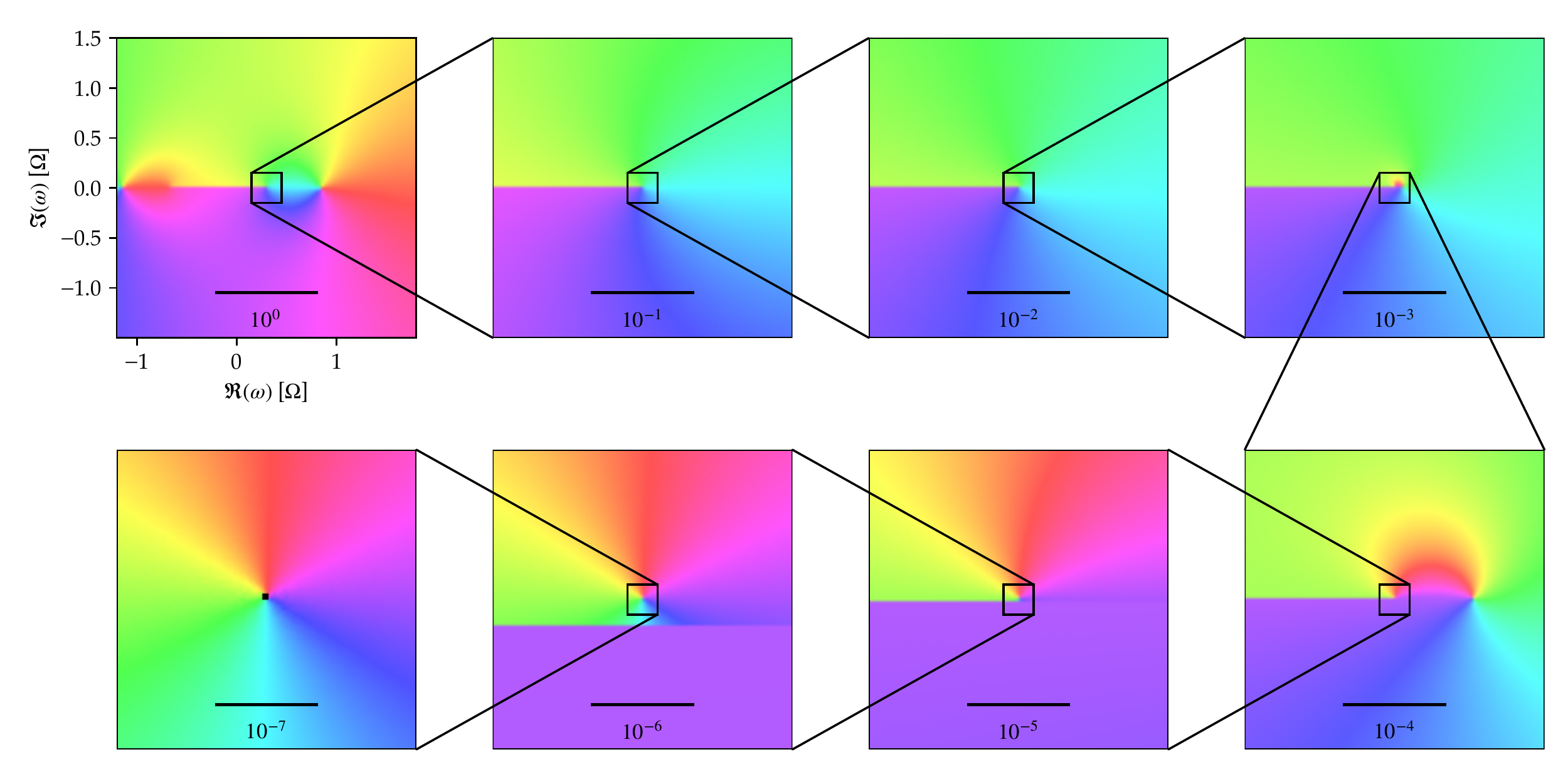}
    \caption{Colourized visualisation (also known as domain colouring) of $f_{\rm disp}$, for $K_x=54.11$, $K_z=300$, $\mu=10$ with an MRN size distribution  $\taus=[10^{-8}, 10^{-1}]\Omega^{-1}$. Colour indicates complex phase; brightness indicates absolute value. Clockwise from top left we zoom in by a factor of 10 in each panel. The black horizontal bar indicates the scale of the plot.}
    \label{fig:zoom}
\end{figure*}

Consider for example the integral
\begin{align}
\Im\left(\int_0^1 \frac{\rmd x}{x - \rmi\epsilon}\right) = \int_0^1 \frac{\epsilon \rmd x}{x^2 + \epsilon^2} = \tan^{-1}\left(\frac{1}{\epsilon}\right),
\label{eq:quadrature}
\end{align}
which is representative of the difficult integrals when $\Im (\omega) = \epsilon$. In the limit $\epsilon \rightarrow 0$, the integral approaches $\pi/2$. Results obtained with \quadpack{} \citep{1983qspa.book.....P}, as implemented in {\tt scipy.integrate.quad} are shown in Figure \ref{fig:quadrature}. For a requested error tolerance of $10^{-8}$ (both absolute and relative) the integration fails for $\epsilon < 4.6\cdot 10^{-5}$, with a warning that the integral may be divergent or slowly convergent. For a more stringent error tolerance of $10^{-12}$ (again both absolute and relative), the integration fails for $\epsilon < 2.8\cdot 10^{-6}$ with additional warnings that the result may be affected by roundoff errors. Note that the integration fails despite the fact that the problematic region is at the edge of the integration domain ($x=0$), which is usually beneficial \citep{1983qspa.book.....P}. 

In order to evaluate the dispersion relation accurately for very small growth rates, we have turned to tanh-sinh quadrature \citep{tanhsinh}. The basic idea is to evaluate an integral
\begin{align}
I = \int_{-1}^1 f(x) dx,
\end{align}
by substituting $x = \tanh((\pi/2)\sinh t)$, and approximate the resulting integral over an infinite interval by the midpoint rule with step size $h$:
\begin{align}
I &= \frac{\pi}{2}\int_{-\infty}^\infty f(x(t))\frac{\cosh t}{\cosh^2\left(\frac{\pi}{2}\sinh t\right)} dt \nonumber\\
&\approx \frac{\pi h}{2} \sum_{i=-\infty}^\infty f(x(hi))\frac{\cosh(hi)}{\cosh^2\left(\frac{\pi}{2}\sinh (hi)\right)}\nonumber\\
&\equiv h \sum_{i=-\infty}^\infty f(x_i) w_i.
\end{align}    
Since the weights $w_i$ go to zero double exponentially for $|i|\rightarrow \infty$ \citep{tanhsinh}, the number of terms needed in the sum should be small. The step size $h$ is reduced until the required error tolerance in reached. It is most efficient to proceed in levels, where at each level the step size is reduced by a factor of 2. In this case, half the terms are already known from the previous level and do not have to be recomputed. Furthermore, given a minimum step size $h_{\rm min}$, all abscissae and weights can be precomputed, which is advantageous if, like in our case, one needs to evaluate many integrals. We have chosen $h_{\rm min}=2^{-12}$, and set the maximum number of terms considered in the summation so that we aim for a precision of $\sim 10^{-15}$:
\begin{align}
x_{i_{\rm max}} &< 1 - 10^{-15},\\
w_{i_{\rm max}} &> 10^{-15}.
\end{align}   
The first limit is to make sure we are not evaluating the integrand \emph{exactly} at the end points, which is necessary in cases where the function diverges at an endpoint but the integral is finite. The result for integral (\ref{eq:quadrature}) with the tanh-sinh quadrature is shown in Figure \ref{fig:quadrature} with the green line. Unlike \quadpack{}, tanh-sinh finds the correct limit of $\pi/2$ in the limit $\epsilon \rightarrow 0$. 

The reason for the qualitative improvement over \quadpack{} results lies in the fact that the almost-divergence at the left endpoint gets mapped onto $-\infty$ where the weights go to zero double exponentially. It is therefore important that any difficult points in the integration domain appear at the endpoints only. Most of the time for an integral appearing in our matrix $\mathsf{M}$ this is not the case, which is why we split up the integration domain into two intervals, separated at the size where $\omega = k_x u_x^0(a)$. This way, we have been able to obtain accurate values for $f_{\rm disp}$ for $\Im(\omega)/\Omega \geq 10^{-8}$.

An example of the intricate structure of the dispersion relation is shown in Figure \ref{fig:zoom}, using $K_x=54.11$, $K_z=300$, $\mu=10$ and an MRN size distribution between Stokes numbers $10^{-8}$ and $10^{-1}$. This particular set of parameters gives rise to a growing mode with\footnote{Note that the number of digits in this frequency is chosen so that the result can be reliably reproduced with different sample points, see section \ref{sec:benchmark}.} $\nu = 0.29767708 + 2.8\cdot 10^{-7}\rmi$. In these domain colouring graphs \citep[see e.g.][]{farris}, a complex number is represented by a colour and a brightness. The colour is determined by the argument of the complex number, represented with a hue following the colour wheel. The brightness is determined by the magnitude of the complex number. Therefore, roots show up as dark points with a colour wheel around them, while poles show up as white points with colour wheels in the opposite direction. In the top left panel, two additional roots can be identified by their colour wheels at $\nu=0.83524314-0.014770598\rmi$ and $\nu = -1.1548451-0.0054118162\rmi$, both of which are damped. Zooming in by a factor of 10 in each subsequent panel, we can for the first time make out the root in the second to last panel, where it is located above a branch cut along the real axis, which can be identified by the jump in colour to purple for $\Im(\nu) < 0$. This branch cut indicates that in general, we can not assume $f_{\rm disp}$ to be well-behaved everywhere.   

\subsubsection{Root finding in the complex plane}

\begin{figure}
	\includegraphics[width=\columnwidth]{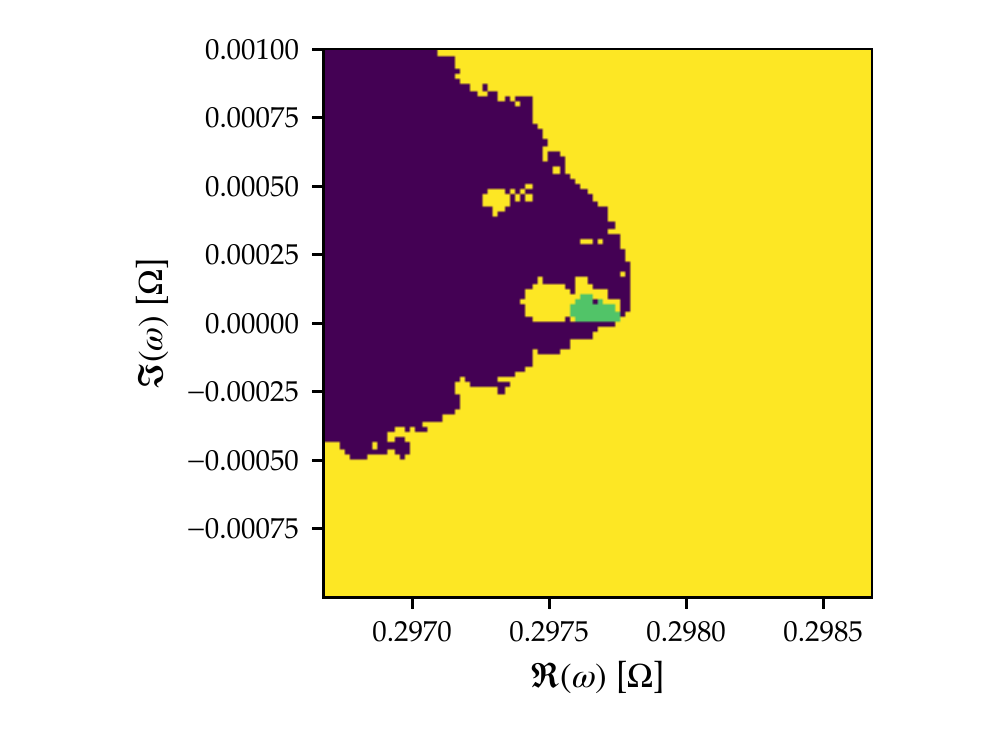}
    \caption{Basins of attraction for the three roots of Figure \ref{fig:zoom}. Dark indicates convergence to the damped mode with negative real part, yellow to the damped root with positive real part, and green convergence to the growing mode.}
    \label{fig:attraction}
\end{figure}

Armed with a reliable algorithm to evaluate $f_{\rm disp}$, the remaining problem is to find its roots for a given set of parameters (in dimensionless form) $(\nu, K_x, K_z, \mu, \sigma^0(a))$. For brevity, we will refer to this problem as finding the roots of $f_{\rm disp}(\nu)$. Note that owing to the complexity of $f_{\rm disp}$ we do not have its derivative with respect to $\nu$ available, which reduces the number of root finding algorithms available. Since $\nu$ is complex in general, the search is two-dimensional, the degrees of freedom being the real and complex parts of $\nu$. While this makes root finding more difficult and expensive compared to root finding in $\mathbb{R}^1$, certain properties of complex numbers can make root finding in the complex plane more straightforward than in $\mathbb{R}^2$, some of which we will detail below. 

Unfortunately, very few numerical root finding algorithms can guarantee to find \emph{all} roots of a function. Iterative methods such as secant or Newton will converge on a root, but the convergence path in more than one dimension is often erratic, and there is often no straightforward relationship between the starting guess and the root the algorithm lands on. For polynomials, this sensitivity of the final answer to the starting guess leads to the concept of Newton fractals, where the boundaries of basins of attraction have a fractal structure. For the same parameters as used in Figure \ref{fig:zoom}, we show the basin of attraction for the growing mode in Figure \ref{fig:attraction} using the secant method. The green region has a size of only $\sim 10^{-4}$ in all directions. This means that without any prior information about the root, we would have to cover a square in the complex plane of size $\sim 1$ (so that both epicyclic and secular modes can be found) with $10^8$ starting points to be able to find this growing mode. This is clearly undoable.  

A very useful technique for finding roots in the complex plane makes use of contour integration. First of all, the number of zeros of an analytic function $f$ inside a contour $C$ can be found from Cauchy's argument principle, which involves a contour integral over $C$ of the logarithmic derivative of $f$:
\begin{align}
Z = \frac{1}{2\pi \rmi} \oint_C \frac{f'(z)}{f(z)}dz,
\label{eq:contour}
\end{align}
where $Z$ is the number of zeros of $f$ inside $C$, counted as many times as its multiplicity. This follows from the fact that $f'/f$ has a simple pole at each zero of $f$ with residue equal to the multiplicity of the zero. A generalization allows the positions of the roots to be calculated \citep{delves}, even without needing the derivative of $f$ \citep{ioakimidis}. This algorithm can in principle guarantee to find \emph{all} roots inside $C$. Unfortunately, our function $f_{\rm disp}$ is not analytic everywhere, as shown for example by the branch cut along the real axis in Figure \ref{fig:zoom}. Even if this can be remedied by choosing $C$ carefully, a more serious issue is the small-scale structure in $f_{\rm disp}$ apparent in Figure \ref{fig:zoom}, resulting in a very large number of function evaluations necessary to calculate the contour integrals. Since evaluating $f_{\rm disp}$ is very expensive, this unfortunately makes this algorithm impractical even if the positions of all branch cuts are known. 
 
A different approach consists of first approximating $f_{\rm disp}$ by a rational function, and using the zeros of the rational approximation as starting points for the secant method. We have opted for the AAA algorithm as described in \cite{AAA}. Working in dimensionless units, given a set of $M$ sample points $\left\{\nu_i\right\}$ with associated data values $\left\{f_{{\rm disp},i}\right\}$, use a subset of $m<M/2$ support points for a rational barycentric interpolation:
\begin{align}
r(\nu) = \left.\sum_{j=1}^m \frac{w_j f_{{\rm disp},j}}{\nu - \nu_j} \middle/ \sum_{j=1}^m\frac{w_j}{\nu-\nu_j}\right. ,
\end{align}
with properties $r(\nu_i) = f_{{\rm disp},i}$ and also $r$ has no poles at $\left\{\nu_i\right\}$. The weights $\left\{w_i\right\}$ are chosen in such a way that $r(\nu)$ is the best fit to the \emph{remaining} sample points in the least-squares sense. If we label the $M-m$ remaining sample points, i.e. those points that are not used as support points, as $\left\{N_i\right\}$ and the associated data values as $\left\{F_i\right\}$, the least squares problem can be cast as 
\begin{align}
\mathrm{minimize} \norm{A w},~\norm{w}=1,
\label{eq:leastsquares}
\end{align}
where $A$ is the $(M-m)\times m$ matrix
\begin{align}
A = \left(\begin{array}{ccc}
\frac{F_1-f_{{\rm disp},1}}{N_1-\nu_1} & \cdots & \frac{F_1-f_{{\rm disp},m}}{N_1-\nu_m}\\
\vdots & \ddots & \vdots\\
\frac{F_{M-m}-f_{{\rm disp},1}}{N_{M-m}-\nu_1} & \cdots & \frac{F_{M-m}-f_{{\rm disp},m}}{N_{M-m}-\nu_m}
\end{array}\right).
\end{align}
The least-squares problem (\ref{eq:leastsquares}) is solved by singular value decomposition \citep{AAA}. If the maximum residual $\max_i |r(\nu_i) - F_i|$ is smaller than a given tolerance (typically set to $10^{-13} \max_i F_i$), the approximation is deemed good enough and the algorithm terminated. Otherwise, an extra support point is chosen from $\left\{N_i\right\}$, followed again by the least-squares problem, until the desired tolerance is reached. The extra support point is chosen to be the one that has the largest residual. Note that in order for the least squares problem to make sense, we need $M-m > m$, or $m < M/2$. In all cases we have considered, $m \ll M$ when the desired tolerance is reached.   

\begin{figure}
	\includegraphics[width=\columnwidth]{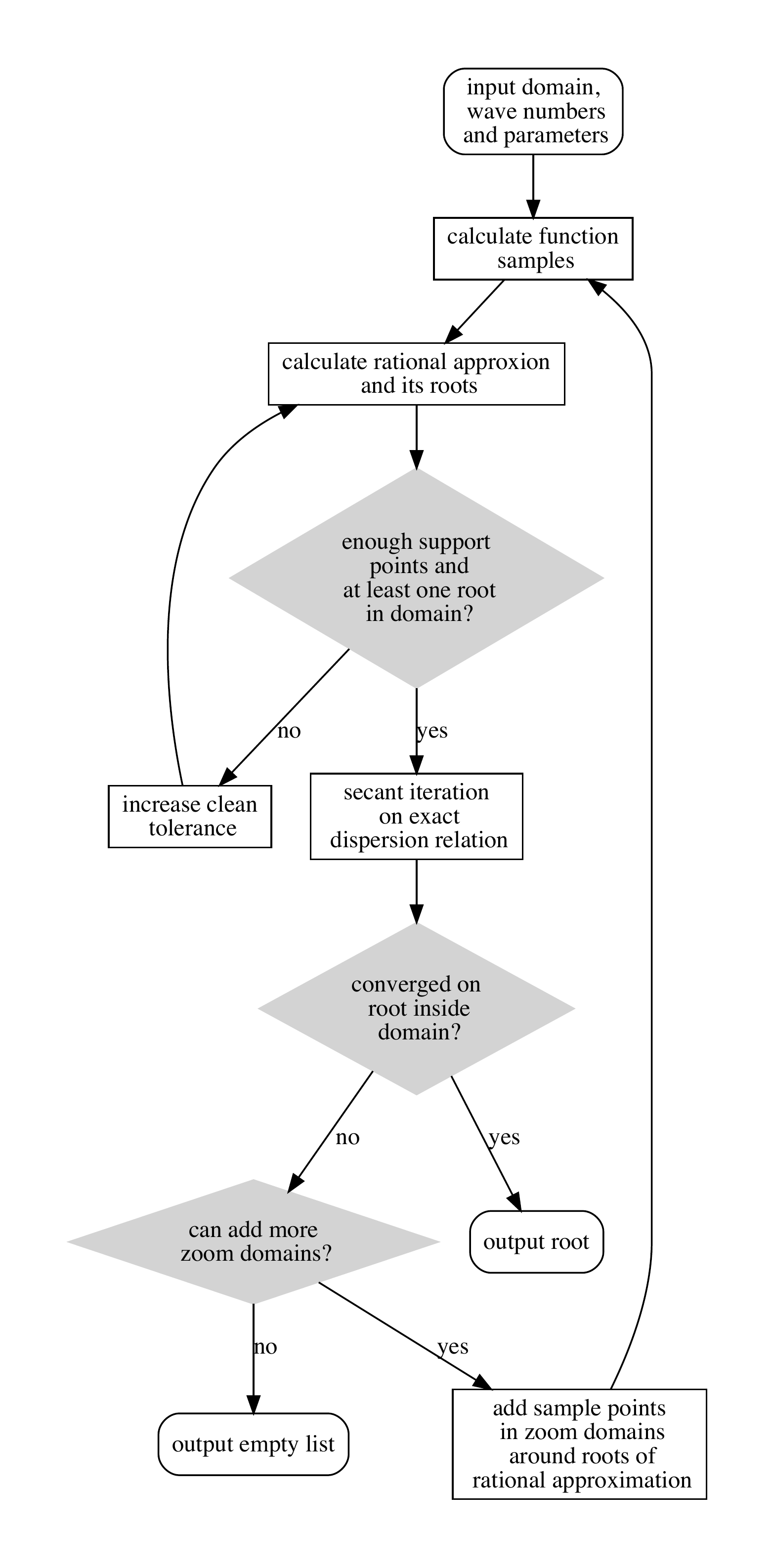}
    \caption{Flowchart of the rootfinding algorithm. Input is the domain $D$, wavenumbers $K_x$ and $K_z$, and the parameters $\mu$ and the size distribution given by $\sigma^0(a)$. Outputs are the roots in $D$, if any.}
    \label{fig:graph_psimode}
\end{figure}

Having obtained a rational approximation $r(\nu)$, its zeros can be found from the generalized eigenvalue problem \citep{klein}:
\begin{align}
\left(\begin{array}{ccccc}
0 & w_1 f_{{\rm disp},1} & w_2 f_{{\rm disp},2} & \cdots & w_m f_{{\rm disp},m}\\
1 & \nu_1 & & &\\
1 & & \nu_2 & & \bigzero\\
\vdots & \bigzero & & \ddots & \\
1 & & & & \nu_m
\end{array}\right) = \nonumber\\
\lambda \left(\begin{array}{ccccc}
0 & & & &\\
 & 1 & & \bigzero &\\
 & & 1 & &\\
 & \bigzero & & \ddots & \\
 & & & & 1
 \end{array}\right).
\end{align}
At least two of the eigenvalues are infinite, and the remaining $m-1$ eigenvalues are the roots of $r(\nu)$. These roots are then used as starting points for secant iterations with the exact dispersion relation. 

The rational approximation often has spurious poles and roots that occur in pairs, the so-called Froissart doublets \citep{froissart, AAA}. Fortunately these are easy to identify since pole and root almost cancel each other in terms of their residue, making the contour integral around the doublet almost zero. In a cleanup step, we therefore take a very small rectangular contour around each root of the rational approximation and calculate the associated residue. If it is smaller than a cleaning tolerance set by the user, the support point closest to this spurious root is removed from the rational approximation. As a result, the spurious root disappears. Care has to be taken with setting the cleaning tolerance: too aggressive cleaning reduces the quality of the rational approximation as many support points are removed. In addition, there is always the chance of accidentally removing a non-spurious root. On the other hand, each root is followed up by a secant iteration on the exact dispersion relation, which is expensive. We decided on a safety net, enforcing a minimum number of support points and at least one root inside the domain of interest. The cleaning tolerance is temporally increased if one of these requirements is not met. 

The rootfinding algorithm takes as input a domain $D$, a connected open subset of the complex plane, which will be searched for roots. Since we are interested in growing modes, we usually take $D$ to be above the real axis. While not strictly necessary for the rootfinding algorithm, it is much more efficient to avoid the branch cut on the real axis. We therefore typically limit ourselves to $\Im(\nu) > 10^{-8}$, and while we then can not exclude growth rates $< 10^{-8}\Omega$, if present these modes would be dynamically unimportant. In practice, it is straightforward, but relatively expensive, to verify that no growing modes exist by letting $D$ drop slightly below the real axis. On the other end we typically limit $D$ to $\Im(\nu) < 1$, since growth rates faster than the dynamical time scale are not expected for the SI. In practice, these modes would be very easy to find (all the difficulties typically occur close to the real axis). It should also be noted that the rational approximation is in principle valid outside $D$ (it only knows about the sample points, not about $D$), and can be quite accurate if the dispersion relation is quite smooth, which is typically the case away from the real axis. This means that it is likely that faster growing modes could be found without enlarging $D$. The real range of $\nu$ we consider is typically $-2 < \Re(\nu) < 2$, so that both epicyclic ($|\Re(\nu)| \sim 1$) and secular ($|\Re(\nu)| \ll 1$) roots can be found. Similar to the fast growing modes, it is likely that any growing modes with $|\Re(\nu)| > 2$ can be found without enlarging $D$.  

We draw sample points from a uniform random distribution over $D$, therefore not including any prior knowledge about any possible roots of $f_{\rm disp}$. For many roots only a few sample points are enough (see below). The root displayed in Figure \ref{fig:zoom}, on the other hand, is one of the most difficult roots to find. In essence, we need a sample point in or very close to the green region in Figure \ref{fig:attraction} in order for the root to be `noticed' by the AAA algorithm. For sample points that are uniformly distributed over $D$, this would mean $\sim 10^8$ sample points, which is again computationally too expensive to be practical. However, we can use the fact that the rational approximation is likely to be good just outside $D$ to make the algorithm adaptive in a straightforward way. 

If we take our standard imaginary range for $D$, i.e. $10^{-8} < \Im(\nu) < 1$, then for any realistic number of sample points the rational approximation does not yield any roots in $D$ for the parameters of Figure \ref{fig:zoom}, or at least the roots do not converge to the growing root upon secant iteration on the exact dispersion relation. However, there are always damped roots of the rational approximation (which are therefore outside $D$). Take such a root $\nu_{\rm damped}$ and create a \emph{zoom domain} $D'$ close to $\nu_{\rm damped}$ but inside $D$. In practice, we take the size of the $D'$ to be $\sim |\Im(\nu_{\rm damped})|$. The zoom domain is populated with the same number of sample points as $D$, and a new rational approximation is calculated based on all sample points of both $D$ and $D'$. This process is repeated until either a growing root is found or the maximum number of zoom levels is reached. 

A flowchart of the basic rootfinding algorithm is shown in Figure \ref{fig:graph_psimode}. The main issue with the algorithm as presented there is that there is no guarantee that \emph{all} roots inside $D$ will be found. In particular, if the rootfinding algorithm returns an empty list, that does not mean that there are no roots inside $D$. It may well be that by increasing the number of sample points and the maximum number of zoom levels a root can be found. If we are interested in only a single set of parameters, it makes sense to throw in the maximum number of points and the maximum number of zoom levels that is computationally feasible. A more common situation, however, is that we are performing parameter scans, for example over wavenumbers. This allows for extra information to be given to the root finder, namely the roots of previously calculated parameters that are nearby. These roots can be passed to the root finder and added as zoom domains automatically at the start. This can speed up parameter scans considerably, making difficult roots much more easy to find, and is exploited by the wavenumber space mapping algorithm discussed in Section~\ref{sec:mapping}.

\begin{figure}
	\includegraphics[width=\columnwidth]{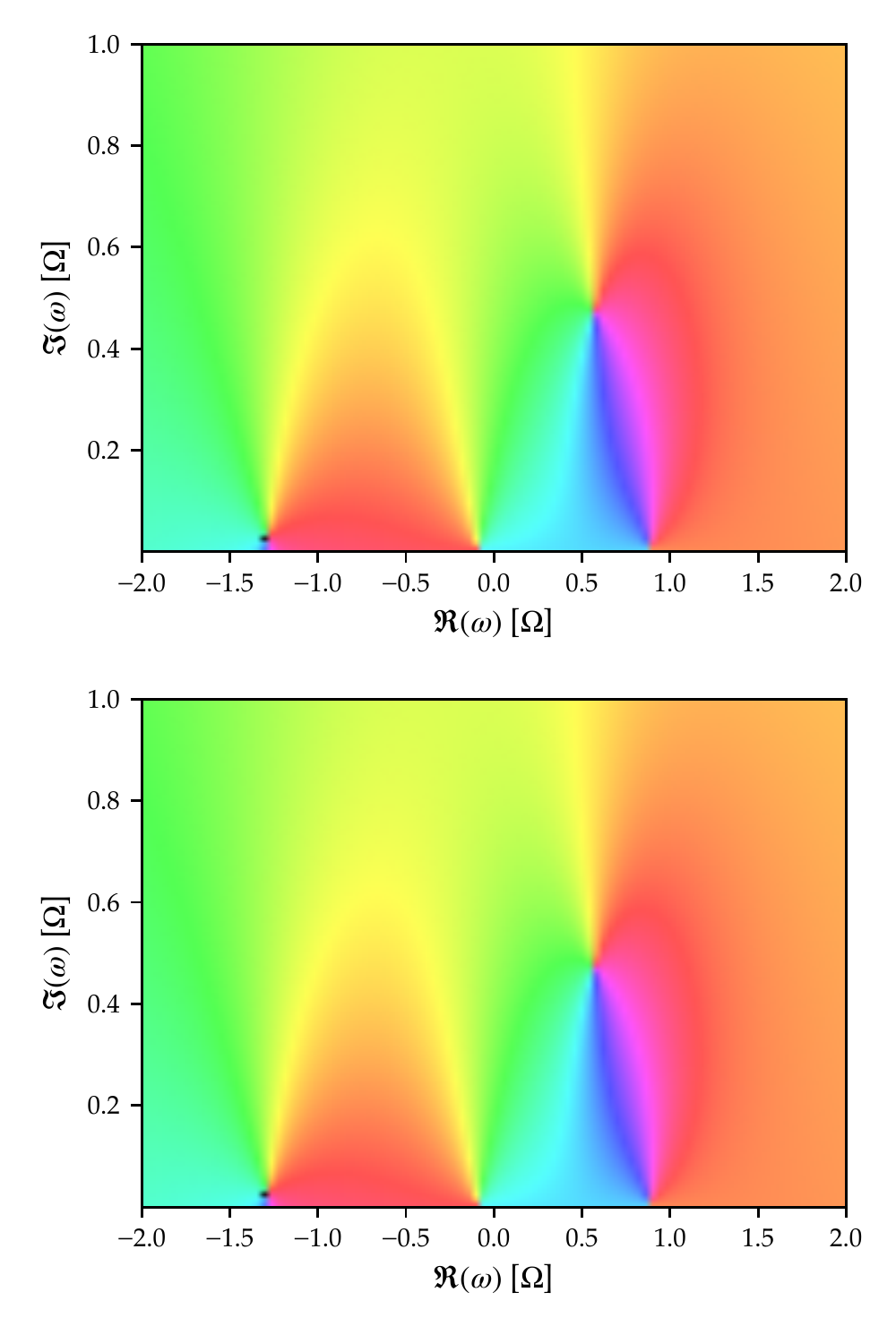}
    \caption{Comparison by way of a domain colouring graph of the exact dispersion relation (top panel) to the rational approximation using $10$ samples (bottom panel) for a single size dust fluid with $K_x=54.11$, $K_z=300$, $\mu=10$ and $\mathrm{St}=0.1$.}
    \label{fig:singlefluid}
\end{figure}

\begin{figure}
	\includegraphics[width=\columnwidth]{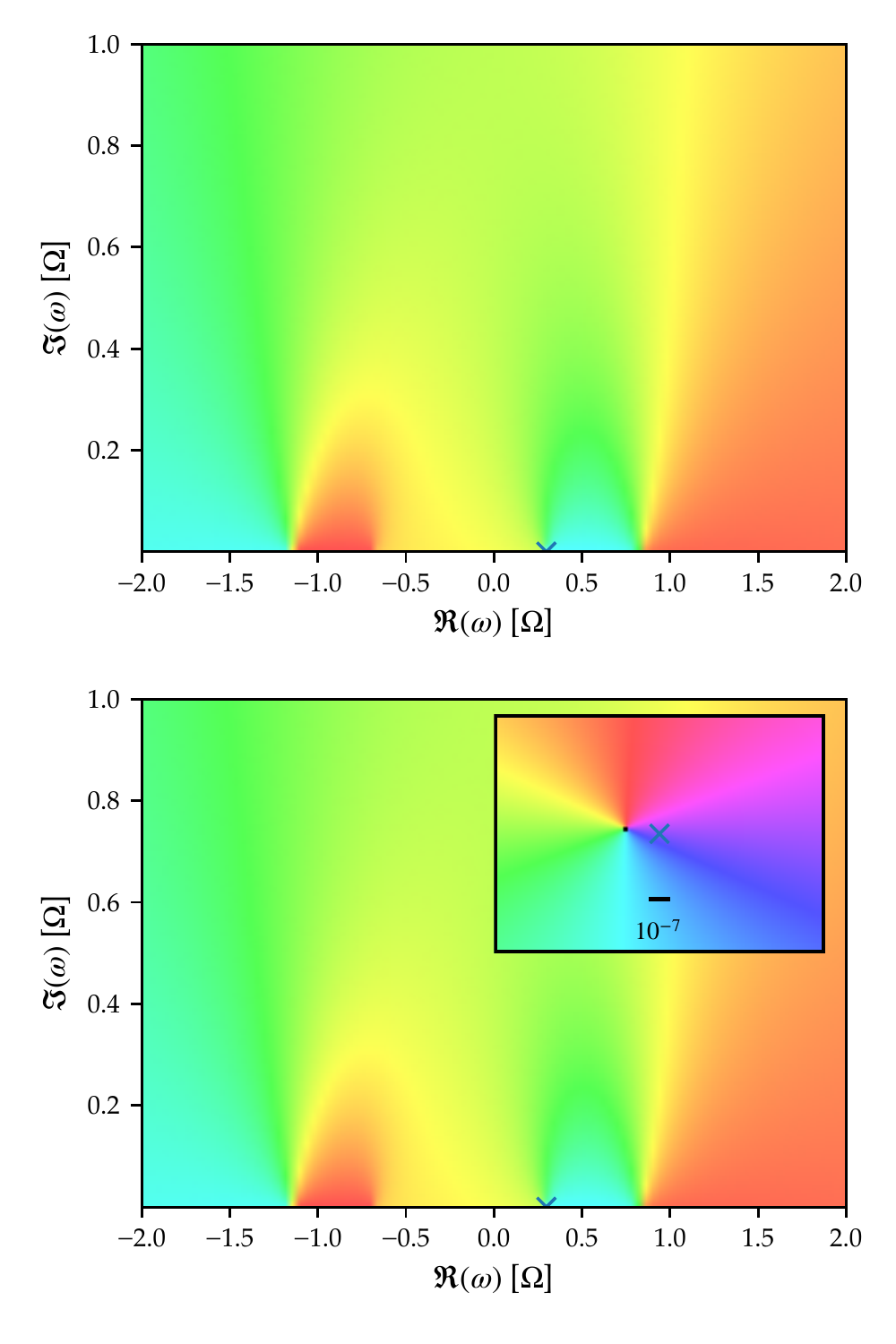}
    \caption{Comparison by way of a domain colouring graph of the exact dispersion relation (top panel) to the rational approximation using $10$ samples and $5$ zoom levels (bottom panel) for the parameters of Figure \ref{fig:zoom}. The inset in the lower panel shows the area around the root of the exact dispersion relation, indicated by the cross.}
    \label{fig:rational}
\end{figure}

\begin{figure}
	\includegraphics[width=\columnwidth]{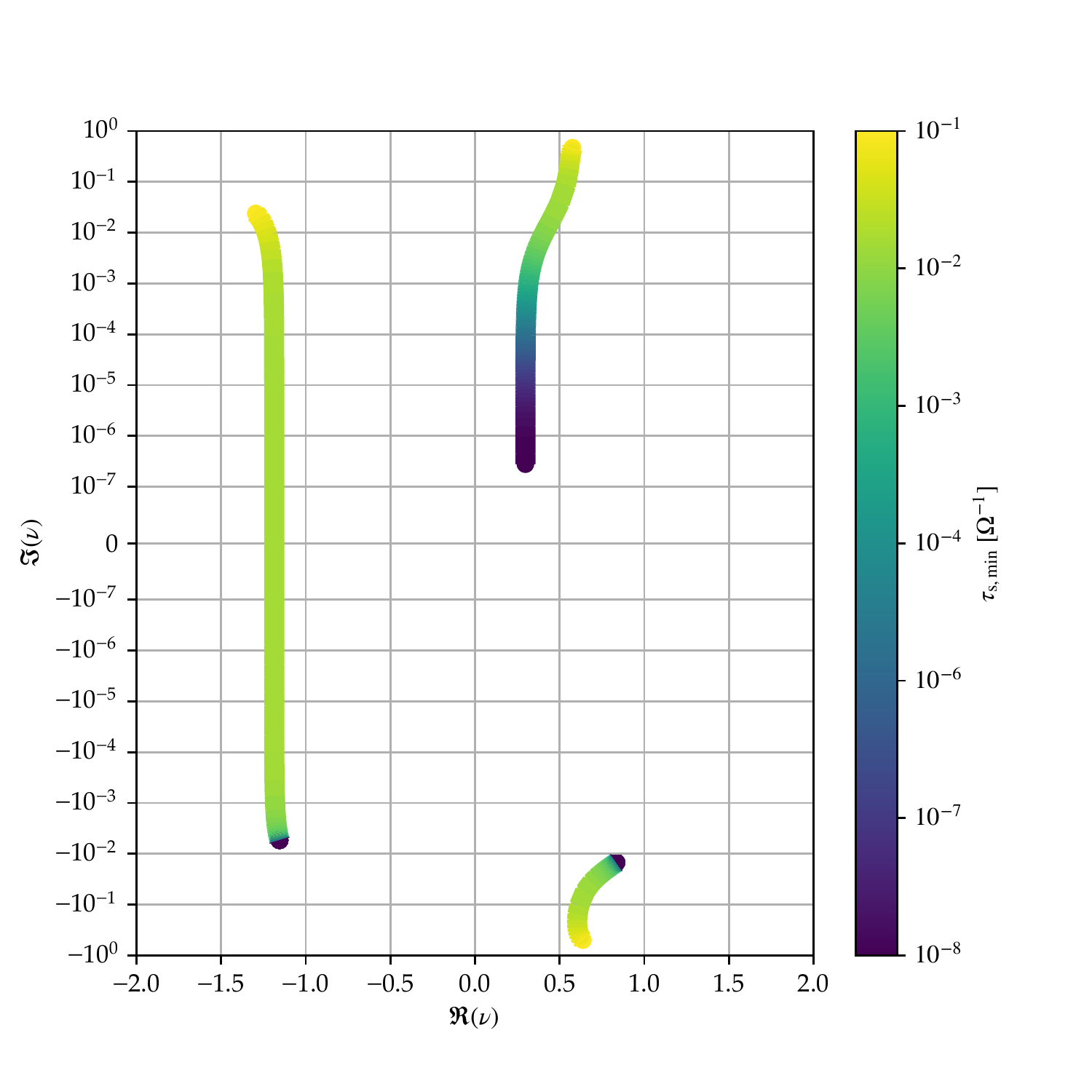}
    \caption{Visualisation of central dispersion relation roots in the complex plane for $K_x=54.11$, $K_z=300$, $\mu=10$ with an MRN size distribution $\taus\in [\tausmin, 10^{-1}\Omega^{-1}]$ with $\tausmin \in [10^{-8},10^{-1}]\Omega^{-1}$, where the colour corresponds to the lower end of the dust size distribution.}
    \label{fig:traject}
\end{figure}

The power of the AAA algorithm is illustrated in Figure~\ref{fig:singlefluid}, where we compare the exact dispersion relation (top panel) to the rational approximation (bottom panel) for the single size limit of the parameters used in Figure \ref{fig:zoom}. The rational approximation was obtained using 10 samples, 5 of which were used as support points. The exact dispersion relation has two growing roots at $\nu=-1.29535+0.023936\rmi$ and $\nu=0.576094 + 0.470201\rmi$, and the rational approximation is almost indistinguishable even for such a small number of samples. The roots of the rational approximation are located at the correct location to 4 significant digits. The sample points $\nu_i$ were drawn from a uniform random distribution with real part $\Re(\nu_i) \in [-2,2]$ and imaginary part $\Im(\nu_i) \in [10^{-8},1]$. While this is one of the easiest roots to find, the accuracy per number of function evaluations (i.e., $10$) is remarkable.   

The parameters of Figure \ref{fig:zoom} give rise to one of the most difficult roots to find. Here the adaptive zoom algorithm is absolutely necessary. Using $10$ sample points per domain, $5$ zoom levels are sufficient to locate the root, giving rise to $170$ sample points in total, of which $39$ support points. The result is displayed in Figure~\ref{fig:rational}. While the global structure is reproduced well without any zoom domains, the fine structure around the root only becomes apparent when using multiple zoom levels. After adding new zoom levels $5$ times, the root in the rational approximation appears close enough to the actual root so that a secant iteration on the exact dispersion relation converges on the growing root. The inset in the bottom panel shows that the root of the rational approximation does not coincide exactly with the root of $f_{\rm disp}$ (indicated by the cross), but a secant iteration gets there in $2$ steps.  

The total number of evaluations of $f_{\rm disp}$ to get the result of Figure~\ref{fig:rational} is $198$. The vast majority comes from calculating the $170$ function samples, while the remaining $28$ are from failed secant iterations after each zoom level. The bulk of the computational time still goes into evaluating $f_{\rm disp}$ rather than the calculation of the rational approximation and its roots. It is worth noting that no prior information about the location of the root was used: the calculation was fully adaptive. 

It is instructive to see how the roots change when the size distribution is widened from single size (Figure~\ref{fig:singlefluid}) to the full width of Figure~\ref{fig:rational}. The paths of the roots are depicted in Figure~\ref{fig:traject}, where we keep the maximum stopping time fixed at $10^{-1}\Omega^{-1}$ and decrease the minimum stopping time from $10^{-1}\Omega^{-1}$ (yellow, single size limit of Figure~\ref{fig:singlefluid}) to $10^{-8}\Omega^{-1}$ (dark blue, the full width of Figure~\ref{fig:rational}). The growing root with negative real part very quickly finds itself below the real axis, at a stopping time range of $\sim 10$. The root that grows fastest in the single size limit progressively moves towards the real axis, never crossing it, and landing on a growth rate of $2.8 \cdot 10^{-7} \Omega$. 

\subsection{Root counting}
\label{sec:rootcounting}

\begin{figure}
	\includegraphics[width=\columnwidth]{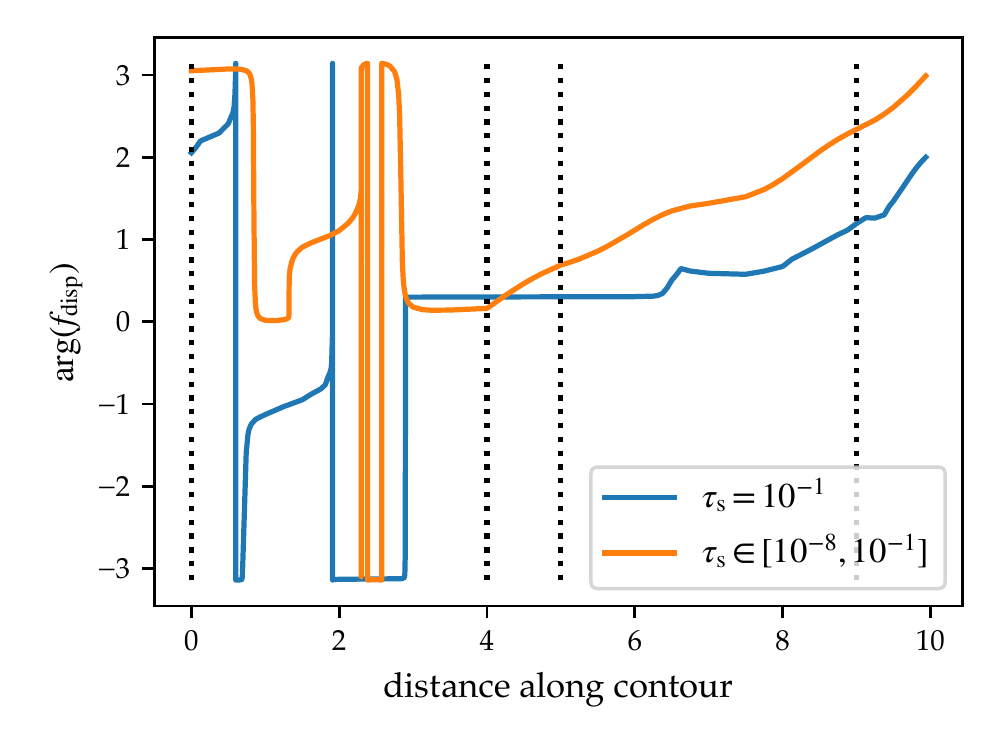}
    \caption{Argument of $f_{\rm disp}$ for the parameters of Figure \ref{fig:singlefluid} (single size, blue curve) and Figure \ref{fig:rational} (size distribution, orange curve). The vertical dotted lines indicate the corners of the contour, from left to right $-2 + 2\cdot 10^{-7}\rmi$, $2 + 2\cdot 10^{-7}\rmi$, $2 + \rmi$ and $-2+\rmi$.}
    \label{fig:arg}
\end{figure}

\begin{figure}
	\includegraphics[width=\columnwidth]{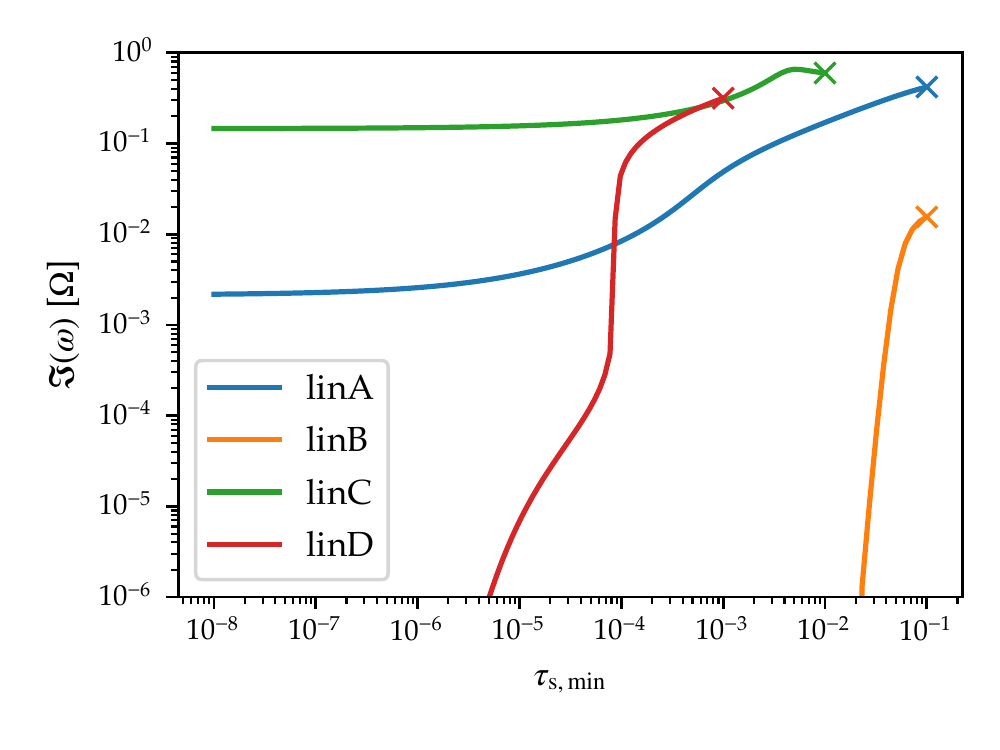}
    \caption{Growth rates of the four monodisperse test cases {\tt linA}, {\tt linB} \citep{2007ApJ...662..613Y} and {\tt linC} and {\tt linD} \citep{2010ApJS..190..297B} when widening the size distribution towards smaller sizes. For each curve, the rightmost end corresponds to the monodisperse limit. Monodisperse results from \citet{2007ApJ...662..613Y} and \citet{2010ApJS..190..297B} are marked by crosses.
    }
    \label{fig:linABCD}
\end{figure}

While finding the roots using contour integral techniques turned out to be impractical, we can still use (\ref{eq:contour}) to calculate the number of zeroes inside a contour $Z$, since it is possible to write the integral in terms of the change in argument of $f$ along $C$. This follows simply from the fact that $\log f = \log (|f|) + \rmi\, \mathrm{arg} f$ (modulo $2\pi\rmi$), and hence
\begin{align}
Z = \frac{1}{2\pi \rmi} \oint_C \frac{\rmd}{\rmd z}\left(\log f(z)\right)dz = \frac{1}{2\pi} \left[\mathrm{arg} f\right]_{z \in C},
\label{eq:argincrease}
\end{align}
where $\left[X\right]_\gamma$ denotes the increase in $X$ along the path $\gamma$. Note that this makes $Z$ the winding number of $f(C)$ around the origin. In practice, one divides up the contour $C$ into small steps, located between a series of points $c_1$, $c_2$, $\dots$. If the argument of $f$ changes by less than $\pi$ between $c_k$ and $c_{k+1}$, the change in argument of $f$ over this step is simply $\mathrm{arg}(f(c_{k+1})/f(c_k))$. Therefore, if this condition is verified for all steps, we have that
\begin{align}
Z = \frac{1}{2\pi} \sum_k \mathrm{arg}\left(\frac{f(c_{k+1})}{f(c_k)}\right).
\label{eq:znum}
\end{align}   
The problem is that without any further knowledge on the global properties of $f$, it is impossible to know whether the argument condition between steps is satisfied \citep[see e.g.][]{yingkatz}. 

In practice this means that it is not possible to get a cast-iron guarantee that the number of roots as found from a numerical implementation of (\ref{eq:znum}) is correct. However, since this is essentially a one-dimensional problem, it is relatively straightforward to implement a reliable refinement algorithm. Starting from the minimum amount of points that define a piecewise linear contour, we keep dividing up segments if either
\begin{itemize} 
\item{The length of the segment is longer than a user-supplied value.}
\item{The change in argument along a segment is larger than a user-supplied limit.}
\end{itemize} 
By default, both parameters are set to $0.1$, which yields reliable results in almost all cases. Looking again at the cases of Figures~\ref{fig:singlefluid} and~\ref{fig:rational}, and defining a rectangular contour with corners $-2 + 2\cdot 10^{-7}\rmi$, $2 + 2\cdot 10^{-7}\rmi$, $2 + \rmi$ and $-2+\rmi$, we find the results depicted in Figure~\ref{fig:arg}. It is immediately apparent that the argument of $f_{\rm disp}$ shows structure on very small scales, so that jumps in $2\pi$, essential for calculating the winding number, can easily be missed. The single size case (blue curve) increases the argument by $4\pi$ before returning to the starting point, indicating there are two roots in the domain, while the full size distribution (orange curve) increases the argument by $2\pi$, indicating that there is only one root inside the domain. These numbers agree with Figure~\ref{fig:traject} in the two limits. To produce the single size curve, $384$ function evaluations were needed, while for the full size distribution $501$ function evaluations were needed. This last number should be compared to the $198$ function evaluations needed to actually find the position of the root using the AAA algorithm. Nevertheless, this technique provides an independent verification that we are not missing any roots.   

\subsection{Wavenumber space mapping algorithm}
\label{sec:mapping}

The root finder algorithm has a remarkable power, but the accuracy and cost is not fully deterministic. This is both due to the role of the random sample points in the AAA algorithm stage and the difficulty of the complex basis of attraction for roots with the secant iteration.
However, we have found that roots with fast growth rates, and those which occur in certain dispersion relation configurations are more easily and reliably located.
We employ an algorithm designed to exploit the strengths and work around the weaknesses of the root finder.

This mapping algorithm is based on using easily found roots as guesses for neighbouring points in wavenumber space.
First, root finder runs are performed on a coarse grid in wavenumber space. This pass most easily picks up the core of islands of growing modes.
Next, another pass is made over the grid, rerunning the root finder algorithm for each point where no growing eigenvalues were found but having neighbouring points with growing eigenvalues, using those neighbour values as guesses which trigger enhanced sampling of the AAA algorithm rational approximation. 
As a protection against excessive work due to redundant roots, the list of guesses is pruned to remove values by dropping the guess $\nu_j$ from the set of guesses when $\|\nu_i-\nu_j\| < 10^{-4}\|\Im(\nu_i)\|$ for $j>i$.
This pass is repeated again if the newly found roots cause any other point with no roots to have new neighbours, until a pass over the grid is made which reveals no new roots. 
Having exhausted all possibilities at the current refinement level, the grid is refined by a factor of two in each direction in wavenumber space, injecting new points in between the existing ones. The cycle of passing over the grid, visiting each point where no root has been found, and using the roots from neighbouring points as guesses for enhanced sampling of the rational approximation is repeated on this refined grid.

As an additional protection against the stochastic nature of the root finding algorithm, for each point in wavenumber space multiple AAA algorithm runs are made with different random number generator seeds. In practice we find four repetitions per point to be sufficient (although conservative).
The cost of repeated solutions can be traded off against a number of other parameters. These include an increased grid resolution yielding more accurate guesses from neighbouring grid points, the use of more samples, or a smaller tolerance in the AAA algorithm.
Tuning the parameters of the wavenumber mapping algorithm affects the cost to find a filled out map, but not the accuracy of the dispersion relation roots found. 
In practice, small omissions in the wavenumber map of fastest growing eigenvalues will not inhibit its interpretation as in many cases the accuracy of the values found is more important

\begin{table*}
\caption{Reference PSI Eigenvalues for an MRN dust distribution}
\label{tab:benchmark}
\begin{tabular}{cccccc}
\hline
$\mu$ & $\beta$ & $\tau_{\rm s}$ $[\Omega^{-1}]$ & $K_x$ & $K_z$ & $\omega \ [\Omega]$ \\
\hline
$0.5$ & $-3.5$ & $[10^{-8}, 10^{-2}]$ & $35$ & $1$ & $0.0238231683+0.002011559\rmi$ \\
$0.5$ & $-3.5$ & $[10^{-8}, 10^{-2}]$ & $500$ & $240$ & $0.3610751+0.01554988\rmi$ \\
$3$ & $-3.5$ & $[10^{-8}, 10^{-1}]$ & $10$ & $0.1$ & $0.00751969859+0.0019668183\rmi$ \\
$3$ & $-3.5$ & $[10^{-8}, 10^{-1}]$ & $80$ & $333$ & $0.66028708+0.12914965\rmi$ \\
\hline
\end{tabular}
\end{table*}

\begin{figure*}
	\includegraphics[width=0.33 \textwidth]{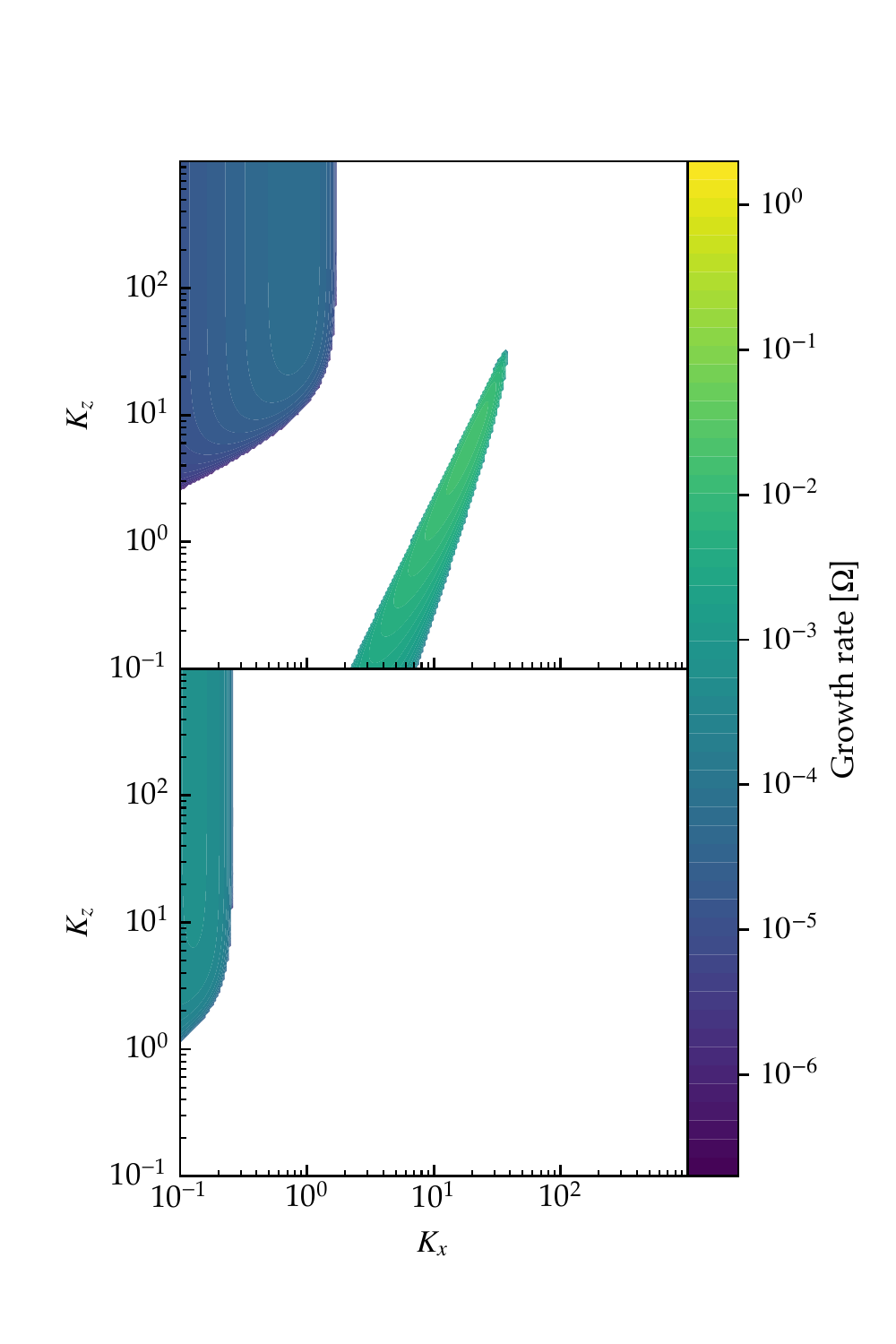}
	\includegraphics[width=0.33 \textwidth]{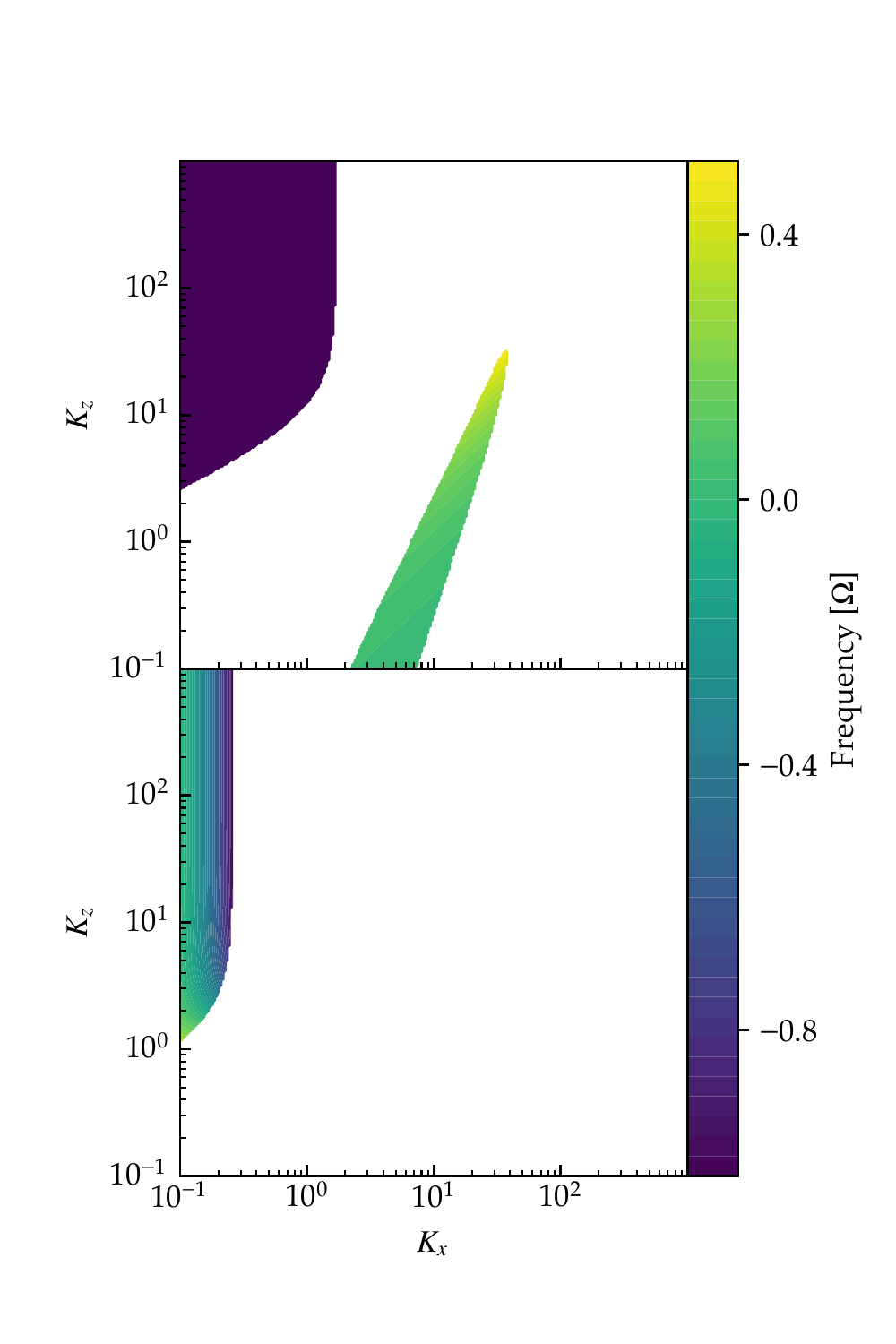}
	\includegraphics[width=0.33 \textwidth]{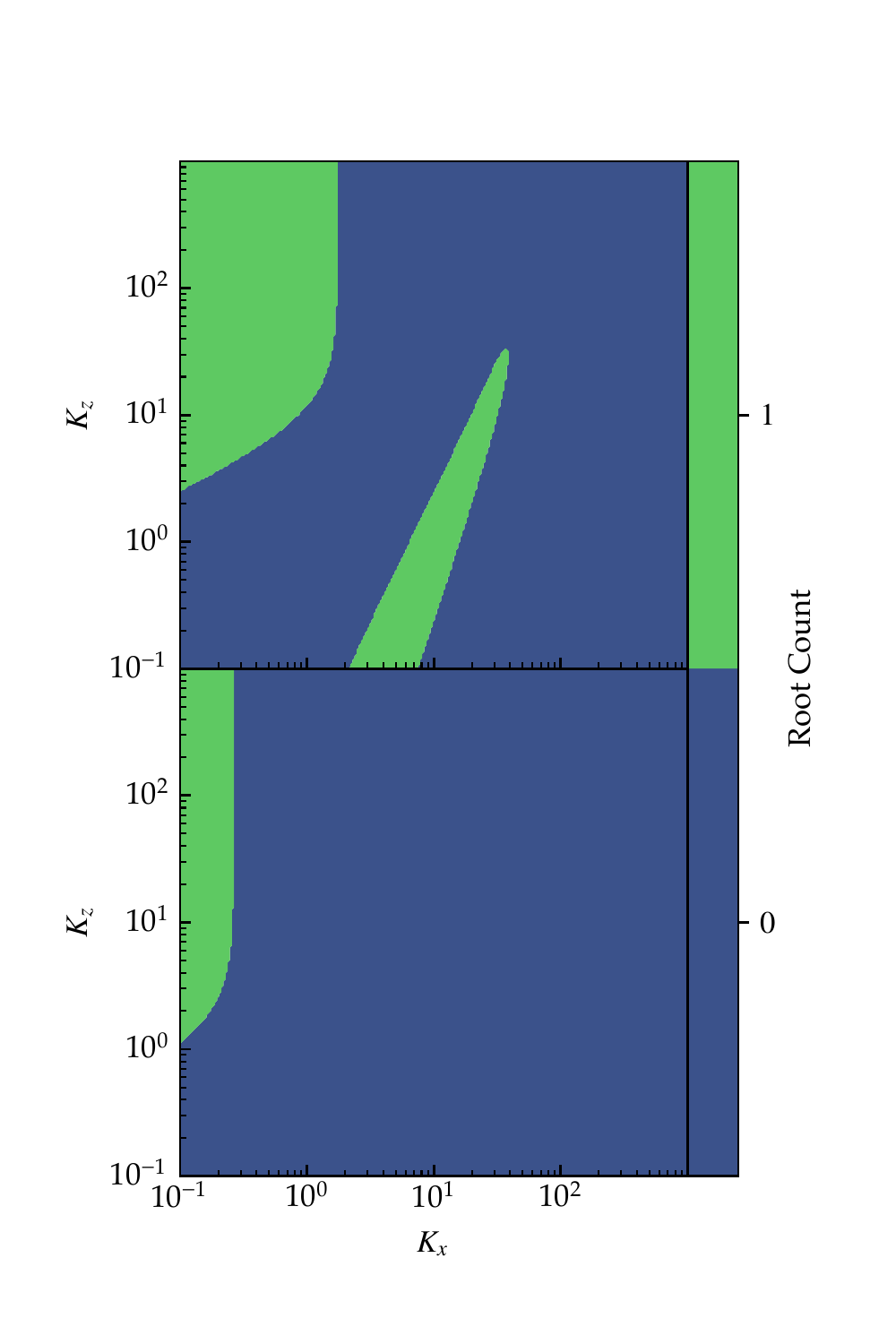}
    \caption{Growing eigenvalues (in units of $\Omega$) for the main cases of \citet{2019ApJ...878L..30K}, with $\mu=1$ and MRN dust distribution.  {\sl Top}: $\taus\in[10^{-4},10^{-1}]\Omega^{-1}$  
    {\sl Bottom}: $\taus\in[10^{-4}, 1]\Omega^{-1}$  {\sl Left}: Growth rate {\sl Middle}: Frequency {\sl Right}: Root count in the domain  $\mathfrak{R}(\omega) \in [-2, 2]$,  $\mathfrak{I}(\omega) \in [2\times 10^{-7}, 1]$.}
    \label{fig:krapp19fig1psigrowth}
\end{figure*}

\begin{table*}
\caption{PSI eigenvalues for \citet{2020arXiv200801119Z} cases}
\label{tab:zy20benchmark}
\begin{tabular}{cccccc}
\hline
$\mu$ & $\beta$ & $\tau_{\rm s}$ $[\Omega^{-1}]$ & $K_x$ & $K_z$ & $\omega\ [\Omega]$ \\
\hline
$2$ & $-3.5$ & $[10^{-3}, 10^{-1}]$ & $60$ & $60$ & $0.50033907+0.09766131\rmi$ \\
$0.2$ & $-3.5$ & $[10^{-3}, 10^{-1}]$ & $10$ & $10$ & No roots  with $\Im(\omega)>2\times 10^{-7}$ exist for $\tausmin < 1.851\times10^{-2}$\\
$0.2$ & $-3.5$ & $[10^{-3}, 2]$ & $1$ & $1$ & $-0.63022118+0.03973764\rmi$ \\
\hline
\end{tabular}
\end{table*}

\begin{figure*}
	\includegraphics[width=
	0.75\textwidth]{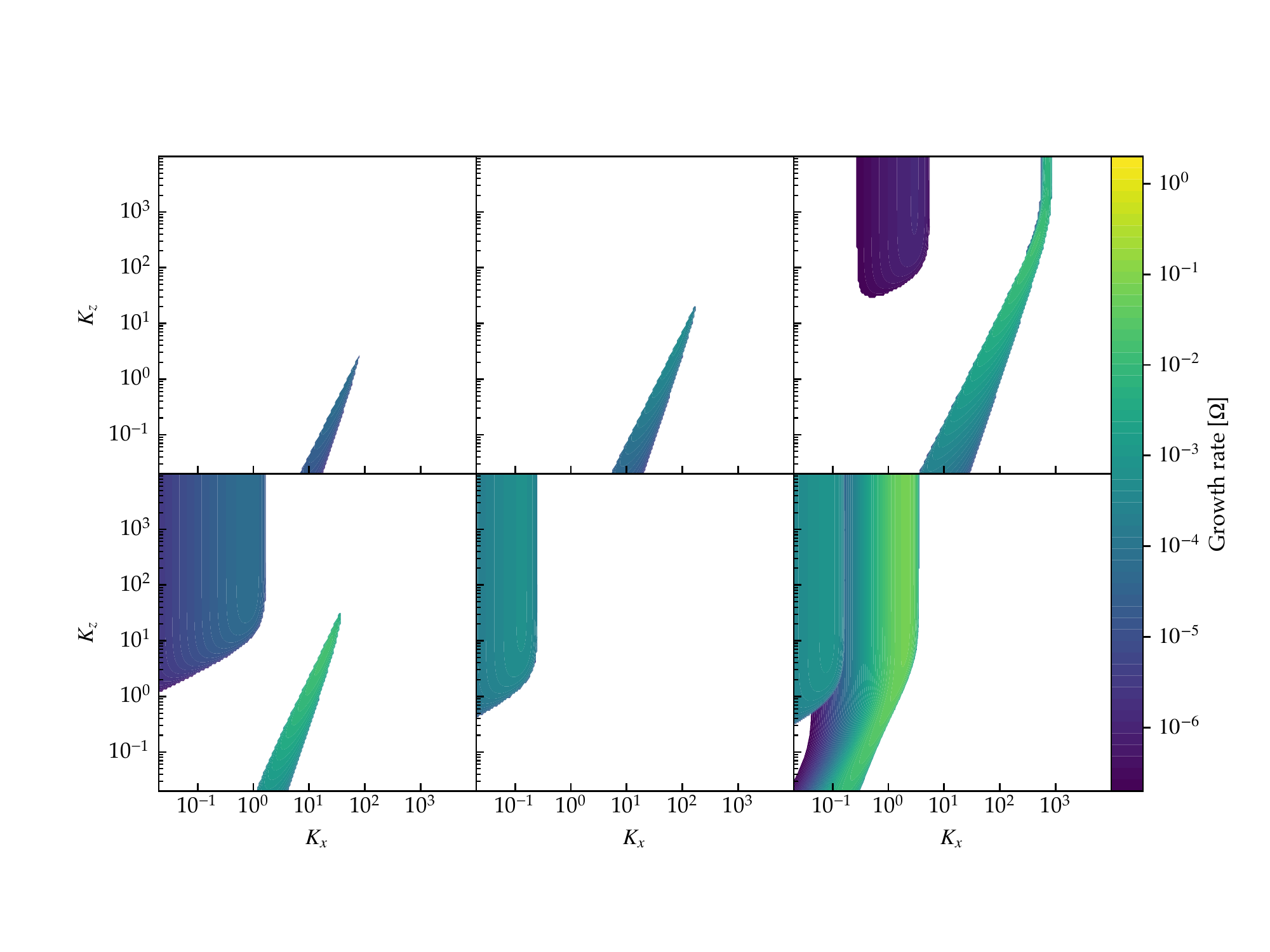}
    \caption{ Growth rate maps for cases from \citet{2020arXiv200801119Z}, all MRN dust distributions. {\sl Top row}: $\taus\in[10^{-4},10^{-2}]\ \Omega^{-1}$ {\sl left} $\mu=0.1$, {\sl middle} $\mu=0.158$, {\sl right} $\mu=0.398$
    {\sl Bottom row}: $\mu=1$, $\tausmin=10^{-4} \ \Omega^{-1}$ {\sl left} $\tausmax=0.1\ \Omega^{-1}$, {\sl middle} $\tausmax=1\ \Omega^{-1}$, {\sl right} $\tausmax=5\ \Omega^{-1}$ .}
    \label{fig:ZY20}
\end{figure*}

\section{Example Computations}

\label{sec:examples}
In this section we use the new algorithms as implemented in {\tt psitools} \citep{psitools}  to calculate some benchmark values and  results for problems previously appearing in the literature \citep{2019ApJ...878L..30K,2020arXiv200801119Z}.
In these examples, the dust distributions are all MRN distributions, which can be specified by a scaling as
\begin{align}
\sigma(a) \propto a^{3+\beta},\ \beta=-3.5\, .
\end{align}
in terms of the dust particle radius $a$.
For all results in this section the radial pressure support of the disc is $\eta=0.05\ r_0\Omega^2$.

\subsection{Monodisperse modes}
\label{sec:mono}

We first consider four benchmark cases that were considered in the monodisperse limit by \cite[][{\tt linA} and {\tt linB}]{2007ApJ...662..613Y} and \cite[][{\tt linC} and {\tt linD}]{2010ApJS..190..297B}. Starting from the monodisperse limit, we increase the width of the size distribution towards smaller sizes using an MRN size distribution, keeping the dust-to-gas ratio constant, until the minimum stopping time is $10^{-8}\Omega^{-1}$. The resulting growth rates are displayed in Figure \ref{fig:linABCD}. At the rightmost point of each curve, which corresponds to the monodisperse limit, we recover the growth rates given in \cite{2007ApJ...662..613Y} to 4 significant digits and those in \cite{2010ApJS..190..297B} to 7 significant digits. The same holds for the real parts of $\omega$. The monodisperse growth rates are indicated by crosses in Figure \ref{fig:linABCD}. When widening the size distribution, two of the modes, {\tt linB} and {\tt linD}, vanish, while the other two reach a limit with non-zero growth, although considerably slower growth than in the monodisperse limit. 

\subsection{Reference PSI Eigenvalues}
\label{sec:benchmark}

To aid in future comparisons and for verifying the functionality of the {\tt psitools} package when run, we have tabulated a set of example roots in Table~\ref{tab:benchmark}.
These have been produced with the root finding technique. To account for the convergence of the final root finding iteration, the solution was repeated from the AAA algorithm with different random sample points ten times, and then result quoted results from truncating the real and imaginary parts to the number of digits having the same value in all ten results. Thus all the digits displayed here should be reproducible and independent of the starting point for the complex root finding iteration.

\subsection{Comparison to Krapp et al.~(2019) cases}
\label{sec:krappfig1}

We present results for the main cases of \citet{2019ApJ...878L..30K} cases in Figure~\ref{fig:krapp19fig1psigrowth} as a demonstration of wavenumber space mapping and root counting.
In these PSI configurations we find the presence of growing modes is isolated to small sections of the  wavenumber space we examined, with no appreciable growth ($>2\times10^{-7}\ \Omega^{-1}$) in most regions.

In the upper panel of Figure~\ref{fig:krapp19fig1psigrowth}, we find two regions of growth above the threshold exist.
The fastest growth is contained in a diagonal ridge, with a growth rate of 
$0.0161\ \Omega$ and frequency $0.232\ \Omega$ at $K_x=20.0$, $K_z=6.56$, similar to the peak value in \citet{2019ApJ...878L..30K} their figure~1.
However, in our results we do not find the ridge of fast growth $\geq 10^{-3}\ \Omega$ continuing to $K_z=10^3$ as \citet{2019ApJ...878L..30K} did with 512 fluids.
Instead, both the root finding calculations and root-counting map do not find any mode with growth above the $2\times10^{-7}\ \Omega$ threshold in this ridge for $K_z > 34.14$.
A second region of growing modes exists along the $K_z$ axis, with growth rates $\sim 10^{-4}\ \Omega$ and frequencies $\approx -1\ \Omega$.
This region also appears in \citet{2019ApJ...878L..30K} their figure~1, with weak dependence on the number of fluids used and a similar growth rate suggesting it is reasonably well converged in their calculations.

Comparing the lower panels of Figure~\ref{fig:krapp19fig1psigrowth} and \citet{2019ApJ...878L..30K} their figure~1 shows the ridge of fastest growing modes disappears in the continuum limit. 
That is, the ridge of growing modes inside the $10^{-3}\ \Omega$ contour in their calculation with 512 dust fluids is completely absent in our results (Figure~\ref{fig:krapp19fig1psigrowth} lower panel) for a continuous dust distribution.
The remaining growth in our continuum calculation is instead along the $K_z$ axis, at a level slightly below $10^{-3}\ \Omega$. This region is in good agreement with the values found in the 512 dust fluid calculation from \citet{2019ApJ...878L..30K}.
Thus, it appears that their 2048 species result on the maximum growth rate of $\sim 10^{-3}\ \Omega$ for this configuration in (\citet{2019ApJ...878L..30K} their figure~4) is very close to the continuum result, but this does not correspond to the region with the fastest growth in their 512 fluid calculation. 

\subsection{Comparison to ZY20 cases}
\label{sec:ZY20}

\citet{2020arXiv200801119Z} have also published results which the {\tt psimode} root finder can be compared to.
First, we provide results for the continuum limit of the cases in \citet{2020arXiv200801119Z} their table~1 in 
Table~\ref{tab:zy20benchmark}.
For the first and last case, the root finding algorithm finds a finite growth rate in good agreement with the highest resolution result given in \citet{2020arXiv200801119Z}.
However, in the $\mu=0.2$, $\taus\in[10^{-3},10^{-1}]\ \Omega^{-1}$, $K_x=10$, $K_z=10$ case we do not find an unstable root of the PSI dispersion relation with a growth rate above $2\times10^{-7}\Omega$ for a dust distribution wider than $\taus\in[1.851\times 10^{-2},10^{-1}]\ \Omega^{-1}$. 
It should be noted that for this case the values given in \citet{2020arXiv200801119Z} are falling nearly linearly with increasing resolution, suggesting agreement with our result that in the continuum case the growth rate is very close to zero.

Results from the root finder and wavenumber mapping algorithm corresponding to  \citet{2020arXiv200801119Z} their figures~8 and~9 are shown in Figure~\ref{fig:ZY20}.
It should be noted that the final one of these cases goes significantly beyond $\taus=1\ \Omega^{-1}$, where in general the fluid formulation for dust is not strictly valid \citep[][see also section \ref{sec:fluid}]{2004ApJ...603..292G, 2011MNRAS.415.3591J}.
However, we have run this case regardless for comparison purposes as the mathematical problem posed is the same regardless of the physical consistency.
Agreement with the better converged regions of the \citet{2020arXiv200801119Z} results is good.
Notably our results also agree that the poorly resolved ridge of apparent growth in \citet{2020arXiv200801119Z} their figure~8 
(Figure~\ref{fig:ZY20} top row), which is rapidly decreasing with resolution and lies along the monodisperse resonant drag instability curve, is not present in the continuum PSI case.
In the cases from \citet{2020arXiv200801119Z} their figure~9 
(Figure~\ref{fig:ZY20} bottom row) we find that two regions of growth for the PSI exist, but unlike \citet{2020arXiv200801119Z} in the $\tausmax=5\Omega^{-1}$ case we find growing modes above our threshold continuously joining both islands of growth.

In discussing their results for these setups, \citet{2020arXiv200801119Z} pose that these leave as a question for future studies of why the PSI shows fast growth at some wavenumbers, and simple direct calculations of the eigenproblem converge only towards very slow or zero growth at others. 
The answer to this is that the PSI size resonance yields instability growth which asymptotically tends to zero as the distribution is widened for many wavenumbers, and only yields finite growth for wide distributions at certain points in wavenumber space, as shown in  \citetalias{paper1}.

\section{Conclusions}
\label{sec:conclusions}
In this second paper of the series, following the presentation of the basic setup and analysis of the terminal-velocity limit of PSI in \citetalias{paper1}, we have presented further mathematical background and advanced numerical techniques for the full PSI problem.
We have presented the first-principles derivation of the governing equations for polydisperse dust-gas flows with dilute dust, the Polydisperse Streaming Instability linear stability problem, and a range of methods for accurately solving the resulting eigenproblem.
We have elucidated the limitations of the a direct methods of solving the eigenproblem based on a discretization of the dust continuum in stopping time as employed by \citet{2019ApJ...878L..30K}, \citet{2020arXiv200801119Z}, and \citetalias{paper1}.
Recognizing these limitations has lead to an alternate formulation of the linear stability eigenproblem 
based on reducing the dispersion relation to a  complex polynomial with accuracy limited only in practice by floating point restrictions.
After describing the structure of this dispersion relation in the complex plane, we have presented an efficient set of algorithms for complex root finding in the specific PSI dispersion relation case.
This circumvents entirely the large computational expense inherent in the previous direct discretizations of the problem, while also providing significant insight with the ability to map the structure of the PSI dispersion relation.

As byproduct, we presented a method for numerically computing the number of roots of the PSI dispersion relation inside a contour in the complex plane through contour integration. This provides another independent check on the convergence and accuracy of the results of the direct or root-finding solvers for the PSI stability problem.

Furthermore, we have presented  a method for efficiently making maps of growing modes in wavenumber space with this root finding algorithm, through recursively refining a mesh in parameter space, and using the results of neighbouring points in phase space to guide the root-finding algorithm, thereby deceasing cost and increasing reliability.

An implementation of all the methods presented is available in the {\tt psitools} package \citep{psitools}. These methods should be useful for general problems involving root finding in the complex plane, in particular when evaluating the function is expensive. Potential examples include finding linear modes in self-gravitating discs \citep[e.g.][]{1989ApJ...347..959A}, and linear modes of the Rossby Wave Instability \citep{1999ApJ...513..805L}.
The next paper in this series will apply these methods to a survey of the most promising regimes where the PSI may provide a mechanism for manifesting the critical planetesimal formation stage in planet formation.

\section*{Acknowledgements}
The software described in this work makes use of the NumPy \citep{2011CSE....13b..22V},
SciPy \citep{2020NatMe..17..261V}, and mpi4py \citep{DALCIN2008655}, pytest, and pytest-mpi libraries.

This research was supported by a STFC Consolidated grants awarded to the QMUL Astronomy Unit 2017--2020 ST/P000592/1 and 2020--2023 ST/T000341/1.
We acknowledge that the results of this research have been achieved using the DECI resource Beskow based in Sweden at PDC with support from the PRACE aisbl.
This research utilised Queen Mary's Apocrita HPC facility, supported by QMUL Research-IT \citep{apocrita}.
SJP is supported by a Royal Society URF.

\section*{Data availability}

The software used to perform calculations in this work is publicly archived on Zenodo \citep{psitools}.




\bibliographystyle{mnras}
\bibliography{psi2} 



\appendix
\section{Fluid approximation}
\label{sec:fluid}

In section \ref{sec:velmom}, the fluid approximation was made for the dust, which amounted to neglecting the stress tensor in (\ref{eq:momstress}). It is worth considering this assumption in a bit more detail for a size distribution. For a continuous size distribution, the most straightforward demand is that the fluid approximation has to be valid across the size spectrum. This means that at any given size, the distribution function has to be sufficiently narrow in velocity space that we can neglect any velocity dispersion, relying on gas drag to keep the velocity dispersion small. Then, if the fluid approximation is valid for the largest size in the size distribution, it will be valid for all sizes. 

In some cases it may be possible to consider larger particles in the size distribution without violating the fluid approximation. Consider the stress tensor appearing in the dust momentum equation:
\begin{align}
\mathsf{T} = \sigma\left< {\bf v}^2\right>_{\bf v}-\sigma {\bf u}{\bf u}.
\end{align}
The evolution of $\mathsf{T}$, integrated over volume and size, is governed by \citep{2011MNRAS.415.3591J}:
\begin{align}
\frac{d}{dt} \left< T^{ij}\right>_{V,a} = 
-2\left<\frac{T^{ij}}{\taus(a)}\right>_{V,a} 
-\epsilon_{\rm ikl}\Omega^l\left<T^{kj}\right>_{V,a} \nonumber\\
- \left<\partial_{x^k} u^i T^{kj}\right>_{V,a} 
- \epsilon_{\rm jkl}\Omega^l\left<T^{ki}\right>_{V,a} 
- \left<\partial_{x^k} u^j T^{ki}\right>_{V,a}.
\end{align}
Here, $\left<\right>_{V,a}$ denotes size-integrated volume average, $\epsilon_{\rm ijk}$ is the Levi-Civita tensor, and ${\bm \Omega}$ is the instantaneous rotation vector of the reference frame. Note this equation is identical to equation (A5) of  \cite{2011MNRAS.415.3591J}, except that in our case we have integrated over dust size as well as volume.

In the single size limit, the evolution of the velocity is dominated by gas drag if the stopping time is smaller than $\Omega^{-1}$ and the viscous heating rate $|\nabla{\bf u}|^{-1}$ \citep{2011MNRAS.415.3591J}. This is essential for making the fluid approximation: it guarantees that whatever the initial state of the system, gas drag will ensure that with time it approaches a fluid-like state with negligible velocity dispersion. In the case of a size distribution, it is the integral over size of $T^{ij}/\taus(a)$ that matters. Noting that $T^{ij} \propto \sigma$, it is clear that if the size distribution is steep enough, it is possible for the high end of the size distribution to have $\Omega \taus > 1$ while the overall system is still dominated by drag. This means that the fluid approximation can still be made, but with a steep size distribution the larger dust particles may not be dynamically important.   

It is important to note that if the initial velocity dispersion is close to zero, the fluid equations will apply for a finite amount of time independent of particle size, provided that there are enough particles to average over. The unstratified, unmagnetised shearing box \citepalias[see e.g.][]{paper1} is perhaps one of the few cases where a dust state with zero velocity dispersion is an equilibrium state. Adding gas turbulence, for example, will unavoidably lead to a velocity dispersion for the dust that will be wider for larger dust sizes. Adding vertical stratification will lead to settling motions for the dust. Therefore, while the unstratified, unmagnetised shearing box serves as a useful simple setup for studying the mSI and PSI, it should be kept in mind that the basic 'fluid-like' state with zero velocity dispersion probably does not apply to more realistic discs.  



\bsp	
\label{lastpage}
\end{document}